\documentclass[draft]{agujournal2019}
\usepackage{url} 
\usepackage[inline]{trackchanges} 
\usepackage{soul}
\usepackage{amsmath}

\usepackage{enumitem}
\usepackage{color}
\usepackage{placeins}
\usepackage{etoolbox}

\draftfalse

\journalname{Journal of Advances in Modeling Earth Systems}

\newcommand\blfootnote[1]{%
  \begingroup
  \renewcommand\thefootnote{}\footnote{#1}%
  \addtocounter{footnote}{-1}%
  \endgroup
}

\begin{document}

%
%


\title{A Spatiotemporal-Aware Climate Model Ensembling Method for Improving Precipitation Predictability}

%
%




\authors{Ming Fan\affil{1}, Dan Lu\affil{1}, Deeksha Rastogi\affil{1}, Eric M. Pierce\affil{2}}

\affiliation{1}{Computational Sciences and Engineering Division, Oak Ridge National Laboratory, Oak Ridge, TN}
\affiliation{2}{Environmental Sciences Division, Oak Ridge National Laboratory, Oak Ridge, TN 37830, USA}




\correspondingauthor{Dan Lu}{lud1@ornl.gov}




\begin{keypoints}
\item We develop a spatiotemporal-aware weighting scheme using Bayesian neural networks for improving model ensemble predictions.
\item The method calculates model skill-consistent weights, provides interpretability, and quantifies uncertainty.
\item We demonstrate the method's superior performance over three baseline ensembling methods in predicting precipitation in CONUS.
\end{keypoints}

\blfootnote{This manuscript has been authored by UT-Battelle LLC, under contract DE-AC05-00OR22725 with the US Department of Energy (DOE). The US government retains and the publisher, by accepting the article for publication, acknowledges that the US government retains a nonexclusive, paid-up, irrevocable, worldwide license to publish or reproduce the published form of this manuscript, or allow others to do so, for US government purposes. DOE will provide public access to these results of federally sponsored research in accordance with the DOE Public Access Plan (http://energy.gov/downloads/doe-public-access-plan).}

%
%

%
%


\begin{abstract}
Multimodel ensembling has been widely used to improve climate model predictions, and the improvement strongly depends on the ensembling scheme. In this work, we propose a Bayesian neural network (BNN) ensembling method, which combines climate models within a Bayesian model averaging framework, to improve the predictive capability of model ensembles. Our proposed BNN approach calculates spatiotemporally varying model weights and biases by leveraging individual models' simulation skill, calibrates the ensemble prediction against observations by considering observation data uncertainty, and quantifies epistemic uncertainty when extrapolating to new conditions.
More importantly, the BNN method provides interpretability about which climate model contributes more to the ensemble prediction at which locations and times. Thus, beyond its predictive capability, the method also brings insights and understanding of the models to guide further model and data development.  
In this study, we apply the BNN weighting scheme to an ensemble of CMIP6 climate models for monthly precipitation prediction over the conterminous United States. In both synthetic and real case studies, we demonstrate that BNN produces predictions of monthly precipitation with higher accuracy than three baseline ensembling methods. BNN can correctly assign a larger weight to the regions and seasons where the individual model fits the observation better. Moreover, its offered interpretability is consistent with our understanding of localized climate model performance. Additionally, BNN shows an increasing uncertainty when the prediction is farther away from the period with constrained data, which appropriately reflects our predictive confidence and trustworthiness of the models in the changing climate.

\end{abstract}


%
%

%


%
%
%
%

\section{Introduction}
Precipitation is one of the key climatic factors affecting fluxes of water, energy, and biogeochemical cycles. It has been observed that climate change non-uniformly shifts regional and seasonal distributions of the precipitation, where dry regions/seasons get drier and wet regions/seasons get wetter \cite{stegall2019}. This shift of precipitation patterns significantly affects natural ecosystem health and human society development \cite{martin2018future, greve2014global}. For instance, in humid regions, the heavy precipitation can increase flood and landslide risks, degrade water quality for human consumption, and disrupt regional ecosystem balance. In arid regions, the decreased precipitation can exacerbate droughts, which leads to water shortages, agricultural production loss, and energy supply risks. Therefore, improving our ability to accurately predict current and future patterns in precipitation is vital for assessing the vulnerability of ecosystems, preparing for extreme precipitation events, and concurrently enhancing water resources management \cite{osti_1761643}. 

Global climate models (GCMs) have been used for improving precipitation prediction and advancing understanding of precipitation's responses to climate change \cite{weigel2021earth,demory2020european}. One of the most inclusive sets of GCMs is from the Coupled Model Intercomparison Project (CMIP), initialized by the Working Group on Coupled Modeling under the organization of the World Climate Research Program \cite{eyring2016overview, taylor2012overview}. CMIP is now in its sixth phase. CMIP6 consists of about 100 GCMs produced by 49 different modeling groups/institutes \cite{zelazowski2018climate}. These GCMs have large uncertainties in physical process representations, show varying prediction skills at different locations and times, and are usually not constrained by observations \cite{eyring2019taking}. Each of these aforementioned factors affects the accurate prediction of precipitation at regional scales. One strategy that can improve the precipitation prediction is a comprehensive multi-model ensembling approach that leverages each individual model's spatiotemporally varying predictive skill, integrates observations to reduce prediction bias, and quantifies predictive uncertainty using a formal calibration and uncertainty quantification (UQ) framework \cite{que2020spatiotemporal,fotheringham2015geographical}. 
 
Several multi-model ensembling methods have been developed. Some approaches assume model independence and model democracy, in which each model is weighted equally. Although studies have demonstrated that under certain conditions equal-weight model averaging could produce better prediction performance than the individual models \cite {gleckler2008performance,knutti2010challenges,pincus2008evaluating}, the assumption on model independence and democracy is not true. Many GCMs in CMIP share components or are variants of other models in the ensemble, and these models have large inconsistency in their skills at a given location and time \cite{alexander2015software,abramowitz2015climate,sanderson2015representative,bishop2013climate}. Even an individual model shows considerably inconsistent skills in different locations and at different times. 
By recognizing the distinct capabilities among the models, some studies assigned unequal weights to individual ensemble members \cite{amos2020projecting,brunner2019quantifying,wenzel2016constraining,karpechko2013improving,raisanen2010weighting}. One of the most frequently adopted ensemble weighting schemes was proposed by  \citeA{sanderson2015representative}. It calculates model weights by balancing the model skill and model uniqueness; the coefficient controlling the balance is determined subjectively, and its value could significantly impact the ensemble results \cite{knutti2017climate,sanderson2017skill}. 

Although some weighted average methods have been proposed, the unequal weights assigned to the individual models are mostly uncalibrated against the observations, an uniform weight is assigned to a model across the space and time, and the same weight is applied for future projections without UQ. Since the model skill varies at regional and seasonal scales, the spatiotemporally uniform weight does not fully leverage each individual model's capability, resulting in the loss of information and possibly large biases in predicting the distribution of precipitation \cite{martin2017connecting,kumar2014regional}.  \citeA{stegall2019} discovered that assigning unequal but spatiotemporally uniform weights to individual models can improve the mean prediction of the precipitation, but the estimated regional precipitation distribution still had a large inconsistency with the observations. 
Additionally, the model weights need to be calibrated against observations in each grid cell at each time step to reasonably reflect the individual model's spatiotemporally varying skill in fitting the observed data and produce observationally constrained ensemble predictions. Studies have shown that many models contributing to CMIP yielded large discrepancies compared with observations, and these model biases should be reduced by calibration before being used for prediction \cite{ukkola2020robust,lorenz2018prospects, mueller2014systematic}. Finally, UQ is required for the ensemble prediction to avoid overconfidence---especially when we project the precipitation in the future changing climate.

In this work, we propose a Bayesian neural network (BNN) ensembling method to improve precipitation predictability by providing accurate and uncertainty-aware predictions. 
The BNN ensembling approach combines GCMs within a Bayesian model averaging framework. It calculates spatiotemporally varying model weights and biases, calibrates the weights and biases against observations, and accounts for the varying quality of the observed data. Additionally, the BNN method quantifies epistemic uncertainty when extrapolating the prediction to new conditions. More importantly, BNN also provides interpretability about each individual model's contribution to the ensemble prediction in different regions and at different times.

The proposed BNN ensembling scheme overcomes the limitations of existing methods by leveraging the power of machine learning (ML) in data analytics and predictive analytics. ML techniques have been applied for predicting precipitation \cite{Jose2022,Heinze2021,Li2021,Ahmed2020}. Most of these applications used ML methods either as a surrogate model of an individual GCM to reduce computational costs in simulation or as a data-driven, black-box regression model to simulate the precipitation directly. The former application considers only a single GCM, and the latter regression model simulation lacks mechanical interpretation and process understanding. Here, we use  ML techniques in the context of multiple model analysis to calculate the model weights of an ensemble of GCMs. The proposed BNN weighting strategy sufficiently leverages each individual GCM's diverse performance in heterogeneous geography and different seasons by calculating spatiotemporally varying model weights and biases. By fusing diverse GCMs, the BNN ensembling embeds our best physical knowledge; and by constraining the ensemble prediction with the observations, BNN enables accurate predictions that match the historical data. Additionally, the BNN method quantifies both aleatoric uncertainty from the data noise and epistemic uncertainty when projecting to the unknown future. Furthermore, besides providing high-quality ensemble predictions with UQ, our method also brings insights and understanding of the climate model performance to guide further model development and prioritize data collection.

We apply the BNN ensembling method for monthly precipitation prediction over the conterminous United States (CONUS). We consider an ensemble of GCMs from CMIP6 and use the European Centre Reanalysis Data (ERA5) as ``observations'' for model calibration and performance evaluation. We perform both synthetic and real case studies to verify, evaluate, and demonstrate the method's capability with respect to prediction accuracy, interpretability, and UQ. Specifically, the main contributions of this effort are as follows.
\vspace{-0.4cm}
\begin{itemize}[leftmargin=15pt]\itemsep0.05cm
\item We propose a BNN ensembling approach for precipitation prediction by leveraging individual GCM's spatiotemporally varying skill and calibrating the model weights and biases against the observations. 
\item We demonstrate the superior prediction performance of the proposed method in comparison with three widely used ensembling approaches on GCMs from CMIP6 and additionally show that BNN can reasonably calculate the epistemic uncertainty in extrapolation to avoid overconfident projections.
\item We investigate the interpretability of the BNN method in terms of which GCMs contribute more to the ensemble prediction at which locations and times and demonstrate that the calculated spatiotemporally varying weights are consistent with the GCMs' simulation skill. 
\end{itemize}

\section{Methods and Data}
In this section, we introduce the BNN ensembling method and describe the climate models and the precipitation data. Next, we briefly introduce three state-of-the-art ensembling schemes with which we compare the BNN for performance evaluation. Lastly, we discuss some evaluation metrics.   

\subsection{Bayesian Neural Networks for Ensemble Model Predictions}\label{sec:BNN}
We assume that observations $y(\mathbf x, t)$ at a given location $\mathbf x$ and time $t$ can be represented as a sum over an ensemble of $m$ GCM predictions $M_i(\mathbf x,t)$ weighted by their respective weights $\alpha_i(\mathbf x,t)$, a bias term $\beta(\mathbf x,t)$, and a data noise term $\epsilon(\mathbf x,t)$: 
\begin{equation}\label{eq:1}
y(\mathbf x, t) = \sum_{i=1}^{m}\alpha_i(\mathbf{x},t)M_i(\mathbf{x},t) + \beta(\mathbf{x},t) + \epsilon(\mathbf x,t).
\end{equation}
The model weights are positive and their sum over the ensemble models is one:  $\alpha_i(\mathbf x,t)>0, ~\text{and}~ \sum_{i=1}^{m}\alpha_i(\mathbf x,t)=1$. The model bias $\beta(\mathbf x,t)$ represents the discrepancy of the weighted ensemble model simulations from the observation. The data noise $\epsilon(\mathbf x,t)$ considers the observation quality varying across the location and time, which is assumed following a Gaussian distribution with a zero mean and a heteroscedastic standard deviation $\sigma(\mathbf x,t)$. The combination of the first two terms at the right-hand side of Eq.~(\ref{eq:1}) forms the BNN ensemble model prediction: $\hat{y}(\mathbf x, t)=\sum_{i=1}^{m}\alpha_i(\mathbf{x},t)M_i(\mathbf{x},t) + \beta(\mathbf{x},t)$. This ensembling scheme expresses the model weights as a function of location and time to leverage individual models' spatiotemporally varying simulation skills. The ensemble prediction additionally considers a bias term that is also a function of space and time. Incorporating the bias term in ensembling is crucial, especially when all the individual models have an over- or under-prediction. In this situation, the weighted ensemble model simulations $\sum_{i=1}^{m}\alpha_i(\mathbf{x},t)M_i(\mathbf{x},t)$ would not perform better than the best-performing individual model, no matter what their weights are. Incorporating the spatiotemporally varying model bias into the ensemble prediction reflects the ensemble model deficiency. 

\begin{figure}[h]
\centering
\includegraphics[width=1\textwidth]{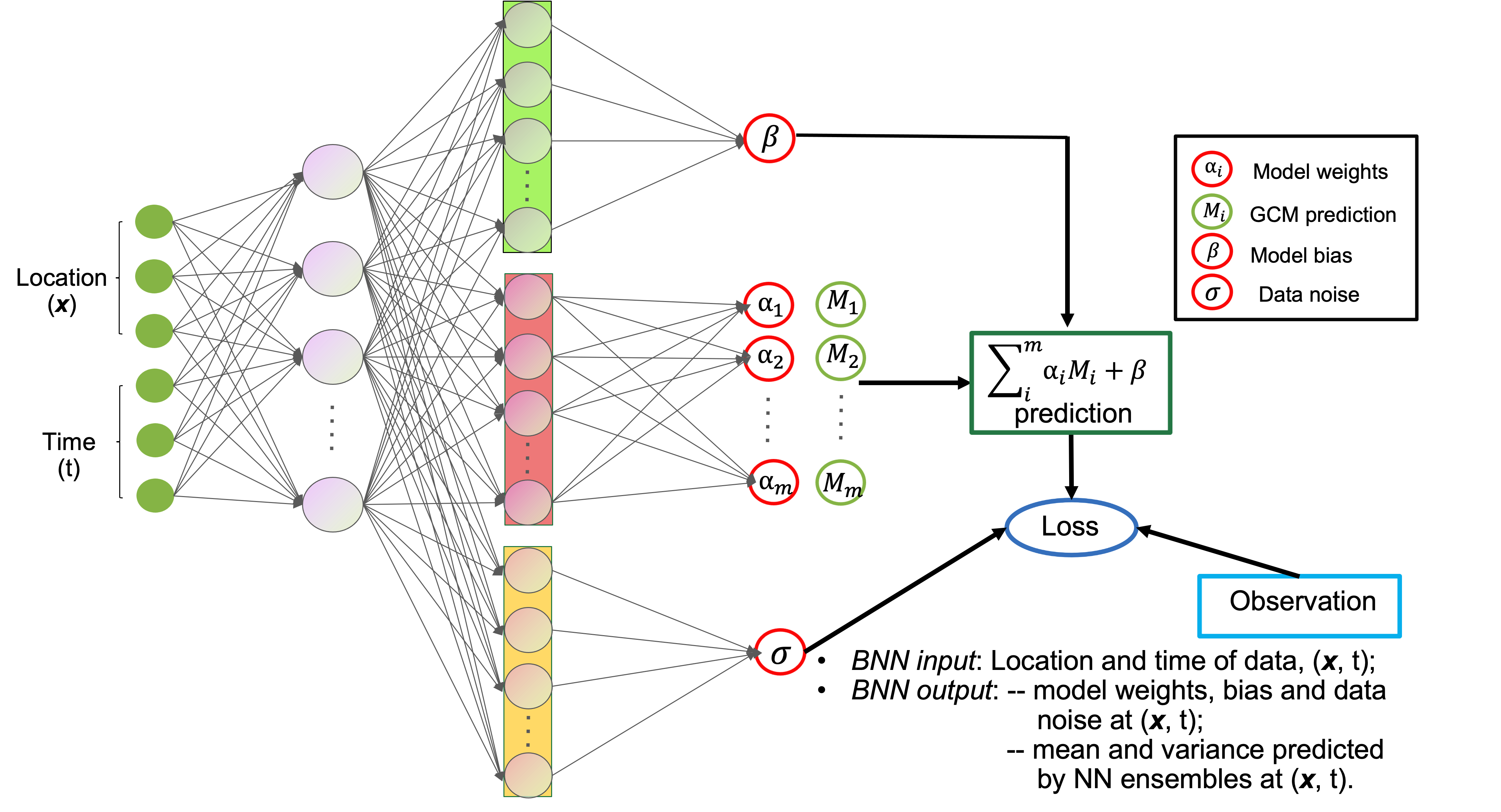}
\vspace{-0.2cm}
\caption{Architecture of the proposed Bayesian neural networks (BNNs).}
\label{fig:archi}
\end{figure}

In implementation, BNN reads the data location ($\mathbf x$) and time ($t$) as inputs and estimates the model weights, biases, and data noises at the given ($\mathbf x,~t$) by calibrating the ensemble prediction against the observations. As illustrated in Figure \ref{fig:archi}, BNN first uses a set of dense layers to extract common information of the model weights, biases, and data noises. Then,  three sets of dense layers are designed  to learn the information specific to each component. Next, BNN incorporates the multiple GCM predictions $M_i(\mathbf x,t)$ and combines them with the estimated model weights, biases, and data noises in the loss function for optimization. The weights, biases, and noises are calibrated as probabilistic functions by specifying distributions over the parameters of the neural networks (NNs) (i.e., we perform the optimization in the Bayesian context). For computational efficiency, we train the BNN using the randomized maximum a posteriori (MAP) sampling \cite{pearce2018uncertainty} instead of the computationally intractable full Bayesian inference, which may require Markov chain Monte Carlo simulation. The MAP sampling approach uses multiple NNs to quantify the ML model parameter uncertainty. Specifically, for the $j$-th network, we draw a sample from the prior distribution over the network parameters (assumed Gaussian) $\theta_{anc,j} \sim N (\mu_{prior}, \Sigma_{prior})$, and compute the MAP estimate corresponding to a prior re-centered at $\theta_{anc,j}$. 
When we consider a dataset of $N$ observations $y_k$ where $k=1,...,N$ and specify the data likelihood by assuming a Gaussian noise with the heteroscedastic standard deviation of $\sigma(\mathbf {x}_k,t_k)$, the calculation of the MAP estimate is equivalent to minimize the following loss function for the $j$-th network: 
\begin{equation}
    Loss_j = \sum_{k=1}^{N}\frac{(y_k-\hat{y}_j(\mathbf{x}_k,t_k))^2}{\sigma_j^2(\mathbf{x}_k,t_k)} + \sum_{k=1}^{N}log(\sigma_j^2(\mathbf{x}_k,t_k)) + ||\Sigma_{prior}^{-1/2} (\theta_j - \theta_{anc,j})||_2^2.
\label{loss function}
\end{equation}
After training, the output of an ensemble of $n_e$ such networks is thus a mixture of $n_e$ Gaussians, $N (\hat{y}_j(\mathbf{x}_k,t_k), \sigma_j^2(\mathbf{x}_k,t_k))$. Then, the mean prediction of these networks $\frac{1}{n_e}\sum_{j}\hat{y}_j$ is the BNN prediction result. The variance $\frac{1}{n_e}\sum_{j}\sigma^2_j + \frac{1}{n_e}\sum_{j}\hat{y}^2_j - (\frac{1}{n_e}\sum_{j}\hat{y}_j)^2$ quantifies predictive uncertainty where the first term quantifies aleatoric data uncertainty, and the combination of the second and third terms quantifies the epistemic uncertainty describing the model's ignorance about the conditions outside the observational records. 

Attributed to this special NN design and Bayesian training, the ensembling strategy of BNN not only calculates spatiotemporally varying model weights and biases, but it also calibrates the weights and biases against observations to fully leverage each individual model's simulation capability, allowing for more accurate and observationally constrained predictions. Furthermore, we trained the BNN using the computationally efficient randomized MAP sampling, which enables rapid quantification of the aleatoric and epistemic uncertainty. Last but not the least, a key strength of this BNN approach is the model’s interpretability, which can explain which models perform well in which locations at which times. This interpretability extends the usage of ML techniques beyond its predictive capabilities to bring insight and understanding to the climate models.

To enable the BNN to produce physically consistent results, we encoded our domain knowledge into the network design and network training. First, in terms of network design, we chose tanh activations for the hidden layers in Figure~\ref{fig:archi} because their mean output is zero-centered, which stabilizes the training. Furthermore, the tanh activations result in a predictably flat extrapolation outside the training set, which ensures a realistic estimation of the model bias and data noise. For the set of dense layers in simulating the model weights, we use a softmax layer at the end to ensure that the model weights sum to unity. 
Additionally, in terms of network training, we first transform latitude, longitude, and time of each data point to a 6 dimensional space-time input. In a climate model, we usually use latitude and longitude to represent a location and use a scalar of $t$ to represent the time (no matter what the unit is). However, directly inputting the three numbers---latitude ($lat$), longitude ($lon$), and time ($t$)---to the BNN would be problematic because the model weights, biases, and data noises generated by such a network would be discontinuous and would not respect seasonality. 
To address this problem, we first represent the location input $\mathbf x$ by its Euclidean coordinate $[cos(lat)sin(lon)$, $cos(lat)cos(lon)$, $sin(lat)]$ and warp the time input $t$ onto a 3D helix $[cos(2\pi t/T)$, $sin(2\pi t/T)$, $t]$, where $T$ is the time scale of the climate model simulation (here $T=1$ month). This transformation of the time variable makes the network generate model weights and biases with both a strong monthly periodicity and a slow variation over the year, which is more consistent with reality. 
Next, we rescale each column of space-time inputs to the range $[-a, a]$ to appropriately represent the varying frequency of the model weights and biases across the space and time. A larger value of $a$ results in a higher changing frequency. In this study, the spatial coordinates are scaled into the $[-2, 2]$ range, and the temporal coordinates are scaled into the $[-1, 1]$ range. 
The network complexity (e.g., the number of layers and the number of nodes in each layer) and the number of networks for Bayesian training are problem specific, depending on the GCM resolution, the model ensemble size, and affordable computing resources. Generally speaking, a large number of complex NNs is needed for an ensemble analysis of many high-resolution GCMs to calculate the spatiotemporally varying weights and quantify the uncertainty, which meanwhile requires a high computational cost. In this study, we use an NN structure in which each set of dense layers in Figure~\ref{fig:archi} has a single hidden layer with 100 nodes, and we use 50 such NNs for UQ. 


\subsection{Precipitation Data and Models}\label{sec:data}
We apply the BNN ensembling method for precipitation prediction based on the GCMs from CMIP6. The simulated precipitation data from the CMIP6-GCMs are downloaded from the Earth System Grid Federation (ESGF) archives (https://esgf-node.llnl.gov/search/cmip6). We consider monthly data from 53 GCMs during the period of 1980--2014, and our analyses focus on the CONUS area. The details of these models are listed in Table~\ref{tab:model}. 
We use the European Centre for Medium-Range Weather Forecasts (ERA5) reanalysis data from the same periods and regions as the reference or ``observations'' for model calibration and performance evaluation \cite{munoz2021era5}. The original ERA5 data are at 33 km horizontal grid spacing and the hourly scale. We aggregate the data to the monthly scale to be consistent with the GCMs simulation data. Both the simulation and reference data are remapped to a common $1^{\circ}$ latitude--longitude grid using the bilinear interpolation method.

\subsection{Three Widely Used Ensembling Schemes} \label{sec:SOTA}
In this section, we introduce three state-of-the-art ensembling schemes, which serve as baselines to evaluate the BNN's prediction performance. The simple average method is straightforward and normally used for multiple model analysis. The weighted average \cite{knutti2017climate} and spatially weighted average methods \cite{amos2020projecting} have an increasing application because of their consideration of model skills and model independence and their good prediction performance. In the following, we briefly describe these three methods where the symbols are consistent with those in Section~\ref{sec:BNN}. \\
\textbf{Simple Average}
The simple average method performs weighted averaging by assigning individual models with equal weights. The ensemble prediction is calculated as
\begin{equation}
    \hat{y}(\mathbf{x},t) = \frac{1}{m}\sum_{i=1}^{m}M_i(\mathbf{x},t) . 
\end{equation}\\
\textbf{Weighted Average}\label{a:weighted}
The weighted average method was introduced by \citeA{knutti2017climate}, who used model ensembles to project the future sea ice change in the Arctic. This weighted average considered model skill and model independence in calculating the weights. For an ensemble of $m$ models, the weight $w_i$ for model $i$ is calculated as
\begin{equation}\label{eq:weighted average}
    w_i = \exp \big(-\frac{D_i^2}{\sigma_D ^2}) / \big(1+ \sum_{j \neq i}^{m} \exp \big(-\frac{S_{ij}^2}{\sigma_S ^2})) ,
\end{equation}
where $D_i^2$ represents the discrepancy between the model $i$ and the observation, and $S_{ij}^2$ describes the difference of the model $i$ from the model $j$. Here, the model outputs and observations in calculation of $D_i^2$ and $S_{ij}^2$ are an averaged value over space and time. This method uses $D_i^2$ and $S_{ij}^2$ to consider model skill and model uniqueness. It also introduces two constants, $\sigma_D$ and $\sigma_S$, to control the influence of the model skill and uniqueness on the weights calculation and, consequently, on the ensemble predictions. For example, when $\sigma_D$ is assigned a small value, only a small number of models obtain weights, whereas when $\sigma_D$ is assigned a large value, this weighted average converges to the simple average with equal weights. Although the values of $\sigma_D$ and $\sigma_S$ significantly affect the ensemble predictions, it is unknown how to assign an appropriate value for a specific problem; currently, the values are determined in a heuristic way. Additionally, although this weighting method considers model skill and independence, it does not consider the model's spatiotemporally varying skill and assigns a uniform weight across space and time.\\  

\textbf{Spatially Weighted Average}
Recognizing that the calculation of $D_i^2$ and $S_{ij}^2$ in Eq.~(\ref{eq:weighted average}) did not consider the difference in space and time, \citeA{amos2020projecting} proposed a spatially weighted average method that calculates $D_i^2$ and $S_{ij}^2$ as a function of location $\bf x$ and time $t$. Specifically, for an ensemble of $m$ models, the spatially weighed average is defined by
\begin{equation}
    w_i = \exp \big(-\frac{D_i^2(\mathbf{x},t)}{n \sigma_D ^2}) / \big(1+ \sum_{j \neq i}^{m} \exp \big(-\frac{S_{ij}^2(\mathbf{x},t)}{n \sigma_S ^2})) ,
\end{equation}
where $n$ is the number of data in calculating $D_i^2(\mathbf{x},t)$ and $S_{ij}^2(\mathbf{x},t)$. Although this method considers model--observation discrepancy and model--model difference across space and time in computing the model weights, it still assigns a uniform weight $w_i$ to an individual model $i$.   
\FloatBarrier
\begin{table*}[h]
\caption{The 53 GCMs from 28 institutes in CMIP6 are considered in this study. The 28 models in bold from each institute are used for ensembling in the real case application in Section~\ref{sec:real}.}\label{tab:model}
\begin{tabular}{ p{1.5cm}   p{7cm} p{5cm}}
\hline
Country  & Research institute  & Model name\\
\hline
Australia   & Commonwealth Scientific and Industrial Research Organization& \textbf{ACCESS-ESM1-5} ACCESS-CM2\\
Canada      & Canadian Centre for Climate Modelling and Analysis & 
 \textbf{CanESM5} CanESM5-CanOE\\
China       & Beijing Climate Center            & \textbf{BCC-ESM1} BCC-CSM2-MR \\
            &Chinese Academy of Meteorological Sciences & \textbf{CAMS-CSM1-0}\\
            &Chinese Academy of Sciences &\textbf{CAS-ESM2-0}\\
            &The State Key Laboratory of Numerical Modeling for LASG& \textbf{FGOALS-g3} FGOALS-f3-L \\
            &Nanjing University&\textbf{NESM3}\\
            & Research Center for Environmental Changes&\textbf{TaiESM1}\\
            &The First Institute of Oceanography, SOA&\textbf{FIO-ESM-2-0}\\
France      &Institut Pierre Simon Laplace&\textbf{IPSL-CM6A-LR}\\
            &Centre National de Recherches Meteorologiques & \textbf{CNRM-CM6-1} CNRM-CM6-1-HR CNRM-ESM2-1\\
Germany     &The Alfred Wegener Institute Helmholtz Centre for Polar and Marine Research  &  \textbf{AWI-ESM-1-1-LR} AWI-CM-1-1-MR\\
           &Max Planck Institute for Meteorology& \textbf{MPI-ESM1-2-LR} MPI-ESM-1-2-HAM MPI-ESM1-2-HR \\
Japan      &The University of Tokyo, National Institute for Environmental Studies, and Japan Agency for Marine-Earth Science and Technology &\textbf{MIROC-ES2L} MIROC6\\    &Meteorological Research Institute &\textbf{MRI-ESM2-0}\\
Italy      &Fondazione Centro Euro-Mediterraneo sui Cambiamenti Climatici & \textbf{CMCC-CM2-HR4} CMCC-CM2-SR5\\
Korea     & Korea Meteorological Administration& \textbf{KACE-1-0-G}\\
          &Seoul National University&\textbf{SAM0-UNICON}\\
Netherlands /Ireland  &EC-EARTH consortium published at Irish Centre for High-End Computing & \textbf{EC-Earth3-Veg-LR} EC-Earth3-Veg EC-Earth3\\
Norway &  Bjerknes Centre for Climate Research, Norwegian Meteorological Institute&   \textbf{NorESM2-MM} NorESM2-LM  NorCPM1\\
Russia& Institute of Numerical Mathematics&\textbf{INM-CM4-8} INM-CM5-0\\
UK&Met Office Hadley Center& \textbf{HadGEM3-GC31-LL} HadGEM3-GC31-MM\\
     &Natural Environment Research Council &\textbf{UKESM1-0-LL}\\
USA         &National Center for Atmospheric Research & \textbf{CESM2-WACCM-FV2} CESM2 CESM2-FV2 CESM2-WACCM  \\
            &Geophysical Fluid Dynamics Laboratory    & \textbf{GFDL-CM4} GFDL-ESM4\\
            &University of Arizona              &\textbf{MCM-UA-1-0}\\
            &NASA/GISS (Goddard Institute for Space Studies) &  \textbf{GISS-E2-1-G} GISS-E2-1-G-CC GISS-E2-1-H\\
            &Department of Energy&\textbf{E3SM-1-0} E3SM-1-1 E3SM-1-1-ECA\\
\hline
\end{tabular}

\end{table*}
\FloatBarrier

\subsection{Evaluation Metrics of Prediction Performance}
We used several statistics and visualization tools to evaluate the prediction performance. For assessing the overall performance, we used root mean square error (RMSE), density plots, and box plots. A better performing ensembling method would have a smaller RMSE value, a closer density/box plot to that of the reference. To evaluate the performance in each grid cell, we present the prediction error across the simulation domain. Additionally, we evaluate the BNN's spatiotemporal-aware weighting scheme by plotting the weights over the spatial domain, in specific regions, and along the simulation time.

\section{Results and Discussions}\label{sec:result} 
To validate and evaluate our proposed BNN ensembling scheme, we applied it to three case studies and compare its prediction performance and weight calculation with the three state-of-the-art methods introduced in Section~\ref{sec:SOTA}. First, we designed a simple numerical experiment in which we know the ground truth to evaluate whether the BNN can accurately calculate the model weights reflecting the individual model's spatiotemporally varying skill. Secondly, we designed a synthetic study where the ``observations'' come from one of the CMIP6 GCMs to further validate the BNN's capability. In the last real case study, we applied the BNN for ensemble precipitation prediction using 28 CMIP6 GCMs from different institutes and use the ERA5 reanalysis data for calibration and evaluation. We analyze the results from three aspects: prediction performance, interpretability, and UQ. 


\subsection{A Simple Numerical Experiment}\label{sec:toy}
\begin{figure}[h]
\centering
\includegraphics[width=1\textwidth]{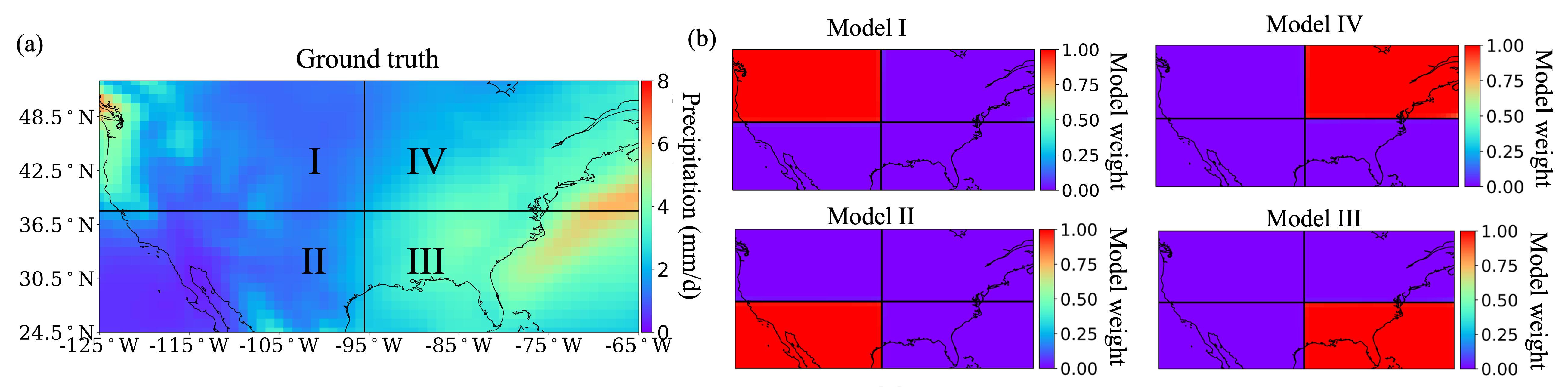}
\vspace{-0.6cm}
\caption{(a) The ERA5 precipitation data used as a ground truth in the numerical experiment, in which we divide the domain into four regions to design four individual models for the ensemble analysis; (b) The BNN ensembling scheme accurately assigns the weight of 1.0 to the regions where the model is accurate and assign the weight of 0.0 to the regions where the model produces random noise.}
\label{fig:S1_weight}
\end{figure}

In the simple numerical experiment, we used 35 years of monthly ERA5 reanalysis data over CONUS in 1980--2014 as the ground truth, based on which we designed four individual models for ensemble analysis. Figure~\ref{fig:S1_weight}(a) shows the averaged ERA5 precipitation data over the 35 years. We divided the simulation domain into four equal regions: region I, II, III, and IV. Model $i$ has the ground truth data in region $i$ (where $i$ represents I, II, III, and IV) and has random noises in the other three regions. We generated the random noise from the uniform distribution in the  $[10, 15]$ range, which is beyond the ground truth, having the maximum average value of $8~mm/d$. We trained the BNN using the first 20 years of data and evaluated its performance on the remaining 15 years.  

\begin{figure}[h]
\centering
\includegraphics[scale=0.45]{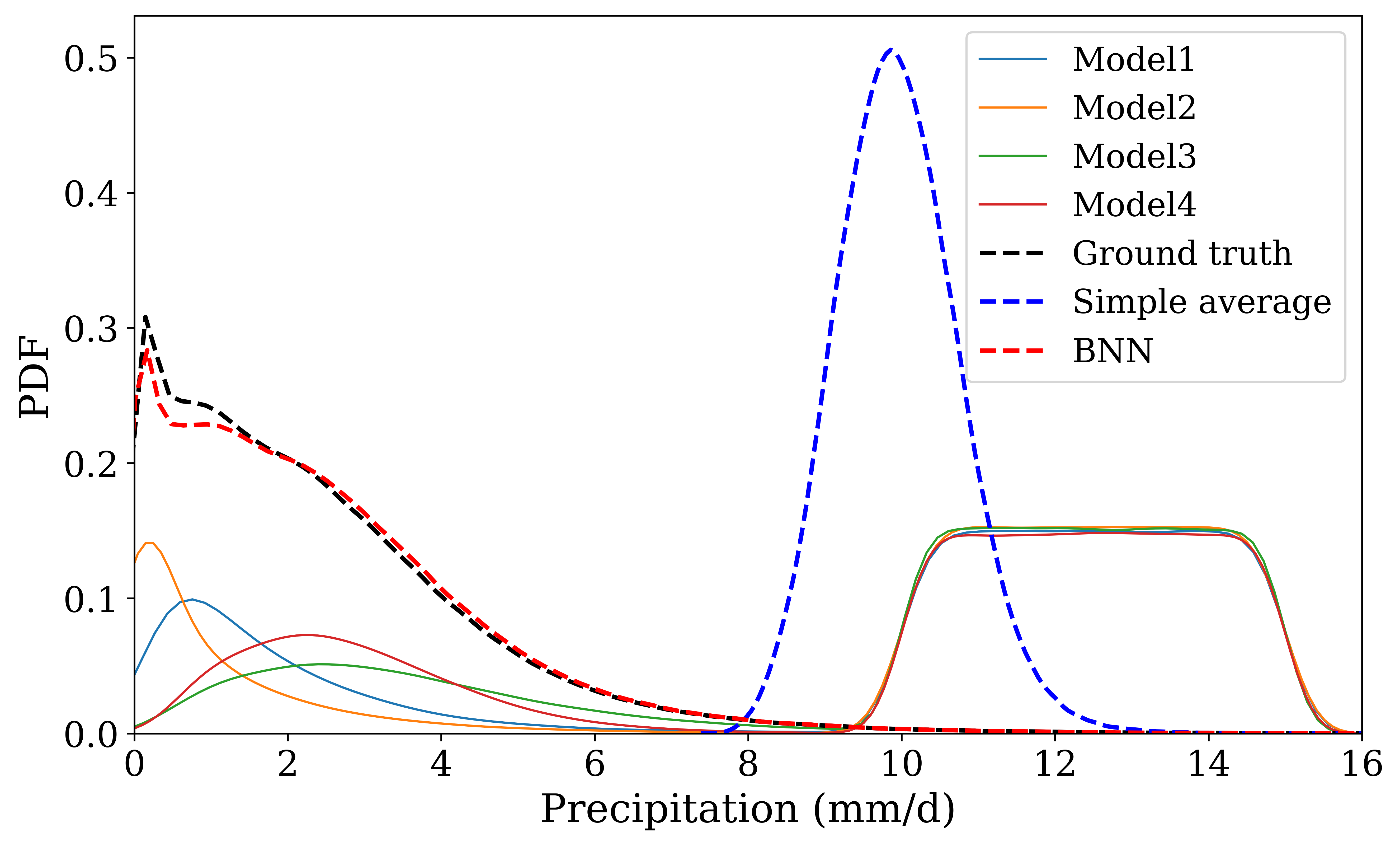}
\vspace{-0.2cm}
\caption{Probability density functions (PDFs) of the precipitation over the entire domain in the out-of-sample test period estimated by the simple average and BNN ensembling approaches, along with the data from the four individual models and the ground truth in the same period.}
\label{fig:S1_precip}
\end{figure}

After training, the BNN calculates the model weights for each grid cell at each month. Figure \ref{fig:S1_weight}(b) summarizes its averaged model weights over the 15 years of the unseen test period for the four individual models in the entire domain. We observed that the BNN successfully recovered the expected model weights; it assigns weights of 1.0 to the regions where the individual model is accurate and weights of 0.0 to those regions where the model produces random noises. Because our BNN reasonably leverages each individual model's prediction skill by accurately calculating the spatially varying weights, its ensemble predictions have a great agreement with the ground truth. As shown in Figure~\ref{fig:S1_precip}, the probability density function (PDF) of the BNN prediction for the out-of-sample test period closely overlaps with the PDF of the ground truth. In contrast, the prediction from the simple average differs dramatically from the truth by assigning equal weights to the models and uniform weights to the entire domain. This numerical example validates this BNN's capability in successfully capturing individual model's spatiotemporally varying skill and demonstrates its competence in accurate ensemble predictions. 

\subsection{A Synthetic Study}\label{sec:synthetic}
\begin{figure}[h]
\centering
\includegraphics[width=1\textwidth]{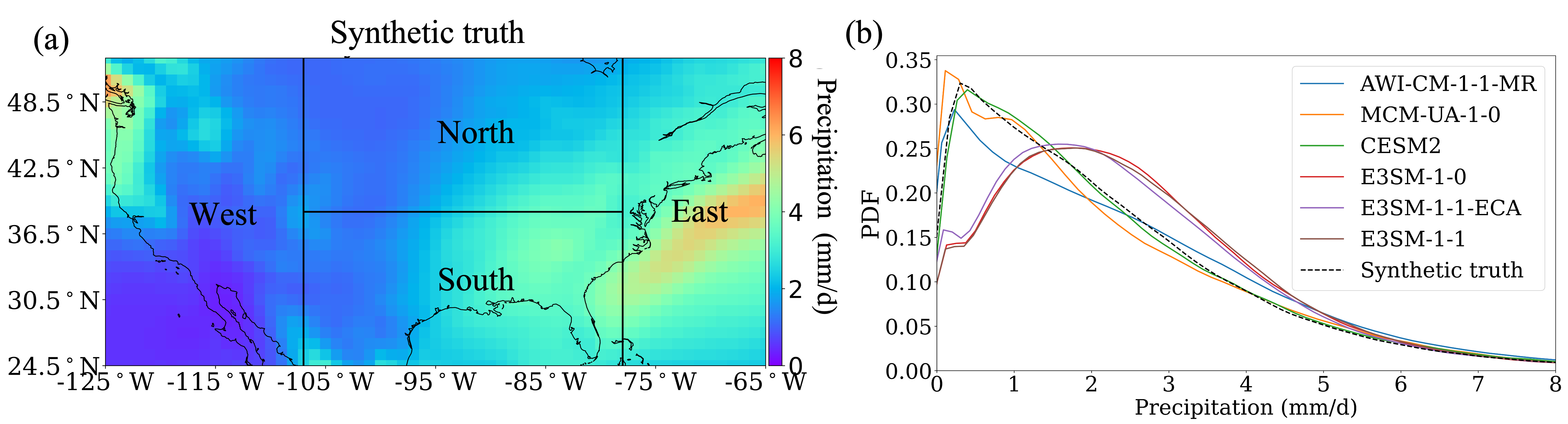}
\vspace{-0.5cm}
\caption{(a) Precipitation data of the synthetic truth averaged over 35 years; (b) The PDFs of the precipitation data from the synthetic truth and the six GCMs for ensemble analysis in the synthetic case study.}
\label{fig:S2_data}
\end{figure}

In this second synthetic case study, we purposely selected seven CMIP6 GCMs from Table~\ref{tab:model} to investigate the BNN's capability. Those seven GCMs are the Alfred Wegener Institute Climate Model (AWI-CM-1-1-MR), Manabe Climate Model v1.0 - University of Arizona (MCM-UA-1-0), 
Community Earth System Model Version 2 (CESM2, CESM2-WACCM), and Energy Exascale Earth System Model (E3SM-1-0, E3SM-1-1, E3SM-1-1-ECA). We chose the simulation data of model CESM2-WACCM as the synthetic truth to calibrate the BNN in the training period and evaluate the BNN's ensemble prediction in the test period. Figure~\ref{fig:S2_data}(a) shows the precipitation data of model CESM2-WACCM in CONUS averaged over the 35 years, and Figure~\ref{fig:S2_data}(b) summarizes the PDFs of the precipitation from the synthetic truth and the six models for ensemble analysis. We can see that models AWI-CM-1-1-MR, MCM-UA-1-0, and CESM2 produce close predictions to the synthetic truth, and the three E3SM models show similar performance, all performing differently from the other four models. In this selection of the individual models and the synthetic truth, we expect that a good-performing ensembling scheme should assign a large weight to those three models, AWI-CM-1-1-MR, MCM-UA-1-0, and CESM2, which produce similar precipitation simulations with the synthetic truth, and assign a small weight to the three E3SM models which have a relatively large discrepancy from the ``truth''. To further investigate the BNN's capability in generating reasonable spatiotemporally varying weights, we divided the simulation domain into four regions---North, East, South, and West (Figure~\ref{fig:S2_data}(a))---to evaluate whether its regional weights reflect the individual model's simulation skill locally. We used the first 20 years of data for training and the last 15 years for out-of-sample testing. 

In the following, we analyze the ensemble prediction results. We first discuss the ensemble prediction accuracy and compare the BNN performance with the three state-of-the-art baselines. Next, we analyze the BNN's weighting scheme in detail by looking at its weights spatially and temporally and investigate the influence of the calculated model biases on prediction performance. In the analysis, we additionally demonstrate the BNN's interpretability. Lastly, we explore the BNN's capability in UQ.  

\vspace{-0.1cm}
\begin{figure}[h]
\centering
\includegraphics[width=1\textwidth]{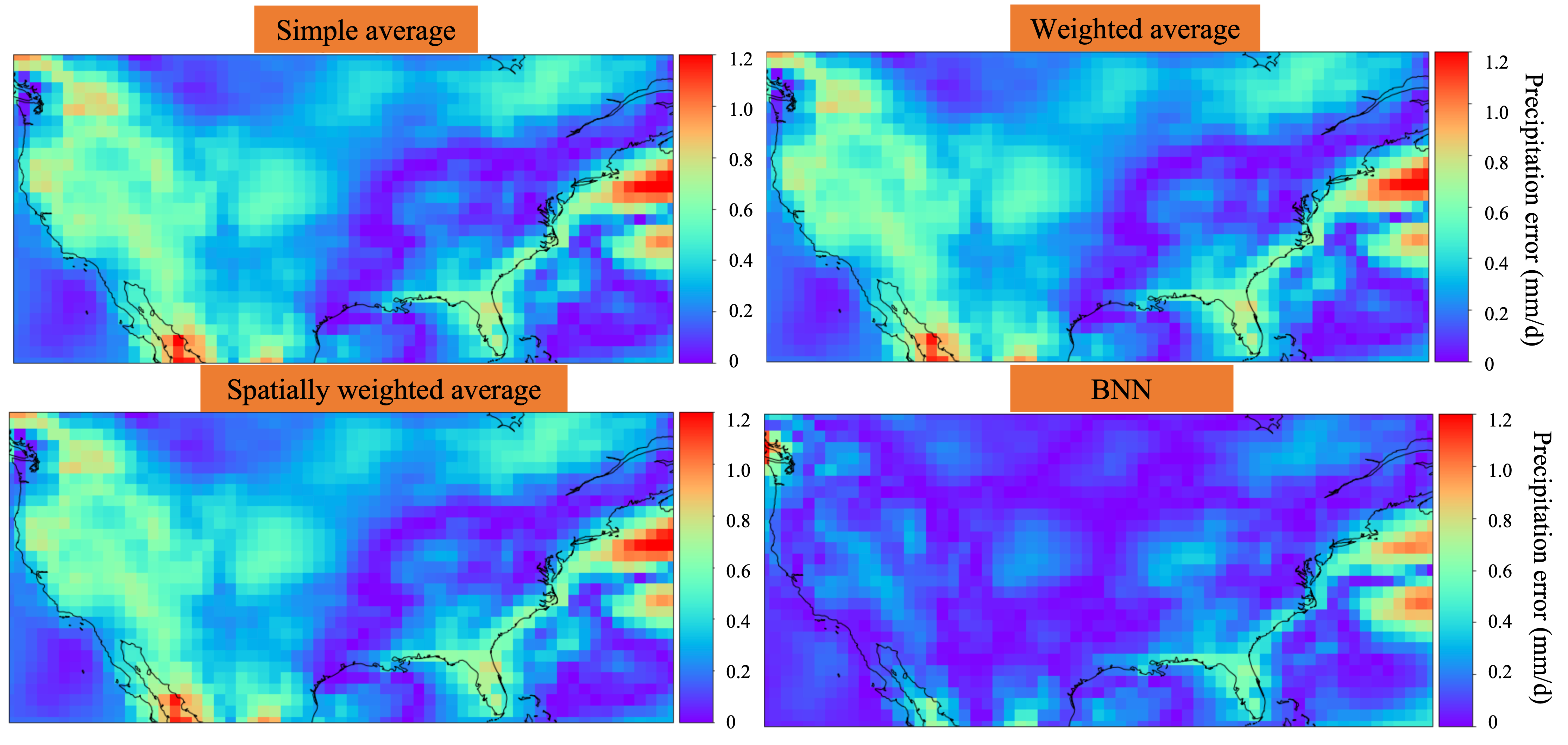}
\vspace{-0.4cm}
\caption{Absolute precipitation errors of the four ensembling methods averaged over the test period in the synthetic study.}
\label{fig:perform1}
\end{figure}

\begin{figure}[h]
\centering
\includegraphics[width=1\textwidth]{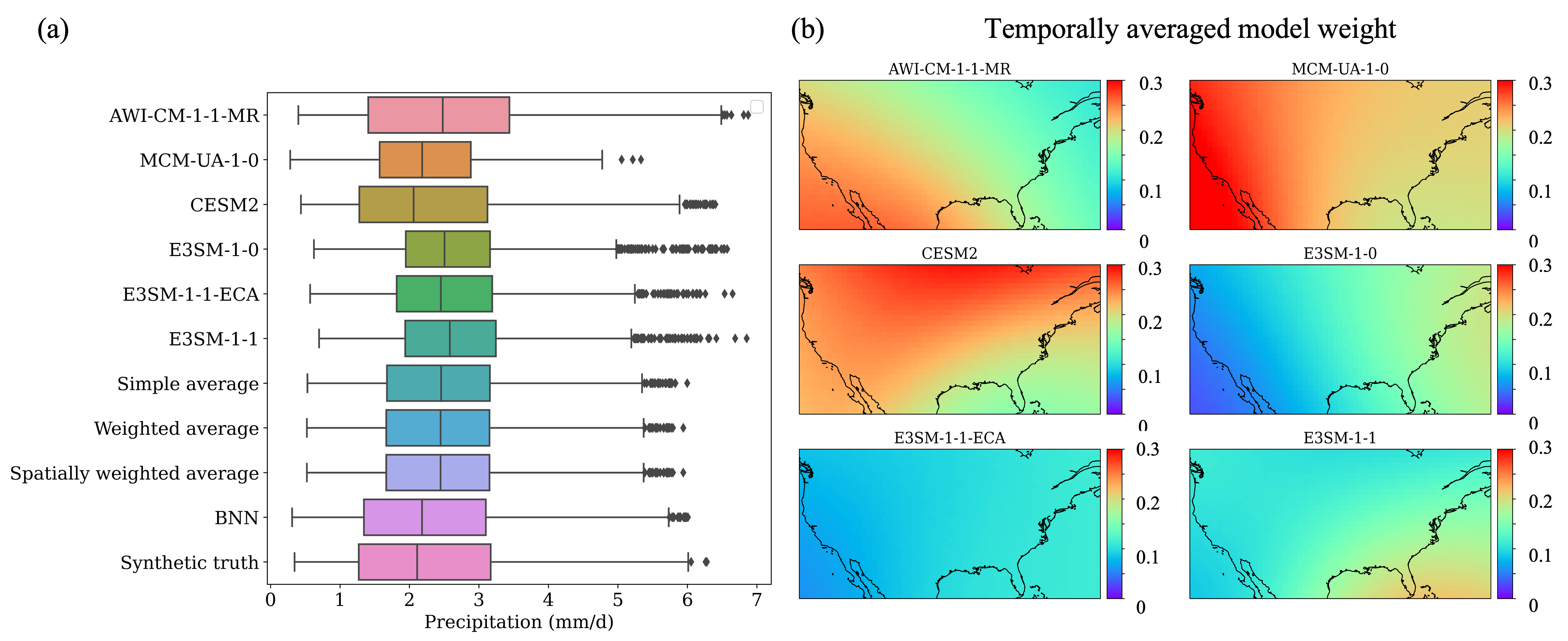}
\vspace{-0.5cm}
\caption{(a) Boxplot of the precipitation data in the test period for the six GCMs and the four ensemble predictions in the synthetic study; (b) Temporally averaged model weights over the test period for the six GCMs.}
\label{fig:perform2}
\end{figure}

Figure~\ref{fig:perform1} shows the absolute prediction errors of the four ensembling approaches averaged over the test period. The figure indicates that BNN produces more accurate ensemble predictions than the other three methods by showing smaller prediction errors in the simulation domain. Figure~\ref{fig:perform2}(a) summarizes the predictions of the six individual models and the four ensembling methods in box plots. The box plots again demonstrate that the ensemble predictions of BNN are closer to the synthetic truth, with similar median and quantiles. On the other hand, the three baseline ensembling methods produce quite similar results, all showing a relatively large difference from the synthetic truth. 
In this case study, we fine-tuned the hyperparameters of $\sigma_D$ and $\sigma_S$ in the weighted average and spatially weighted average methods and show here the best prediction results we obtained after fine-tuning. However, the resulting ensemble predictions from these two weighting schemes do not seem to bring much improvement from the simple average. Their RMSEs are close to each other, with values of 1.44, 1.44 and 1.45 for simple average, weighed average and spatially weighted average method, respectively.
Due to the computational costs, we do not perform hyperparameter tuning for the BNN in this work. However, the current BNN architecture and the set of hyperparameters already show a great improvement in prediction accuracy compared to the three ensembling baselines and the individual models. A higher improvement of BNN is expected after its hyperparameter tuning and architecture optimization. 

\begin{figure}[hbt!]
\centering
\includegraphics[width=1\textwidth]{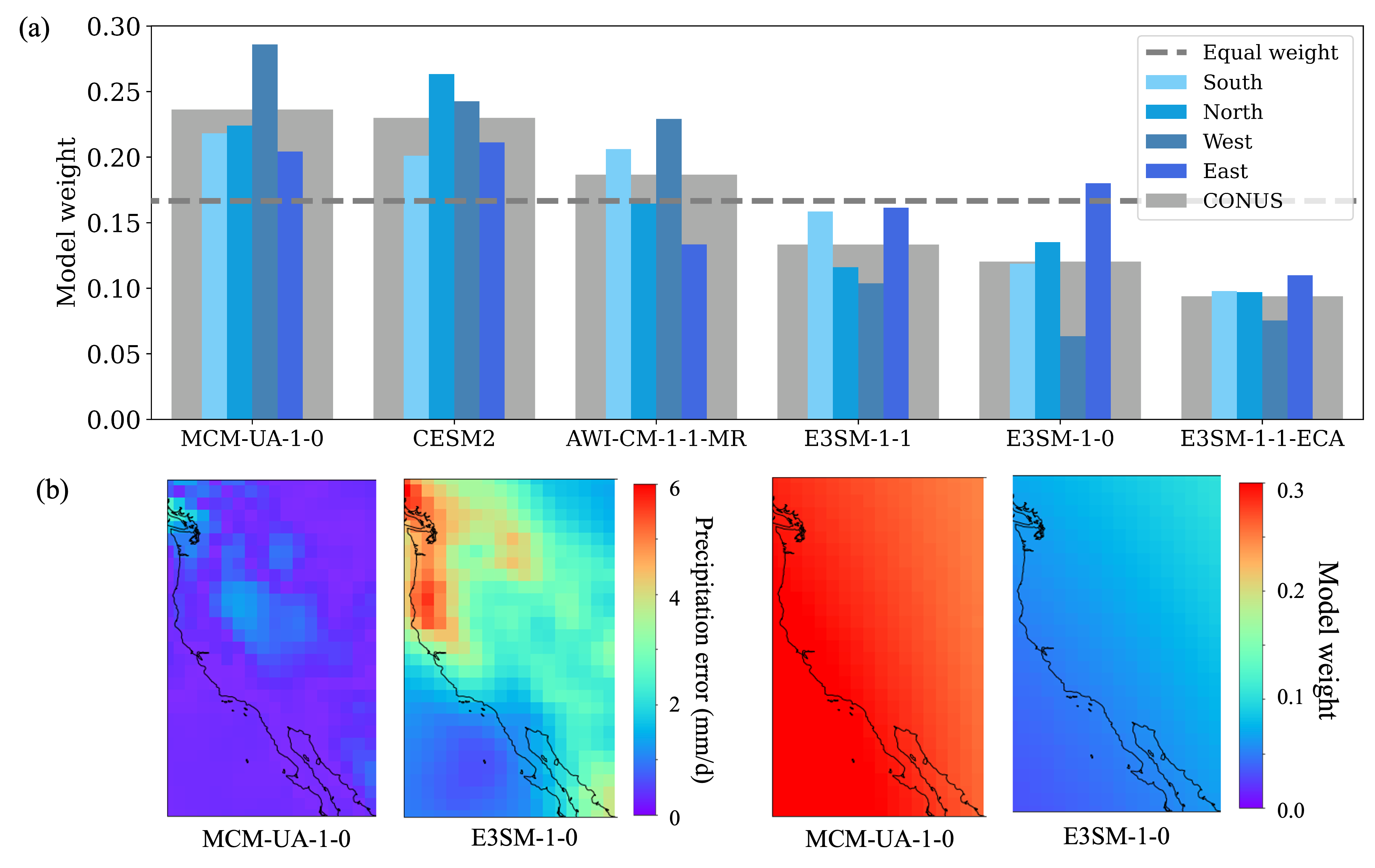}
\vspace{-0.4cm}
\caption{(a) Temporally averaged model weights over the test period in CONUS and the four sub-regions (see Figure~\ref{fig:S2_data}(a)) for the six GCMs in the synthetic study; (b) Prediction errors and model weights of model MCM-UA-1-0 and E3SM-1-0 in the West region.}
\label{fig:spatial}
\end{figure}

\begin{figure}[hbt!]
\centering
\includegraphics[width=1\textwidth]{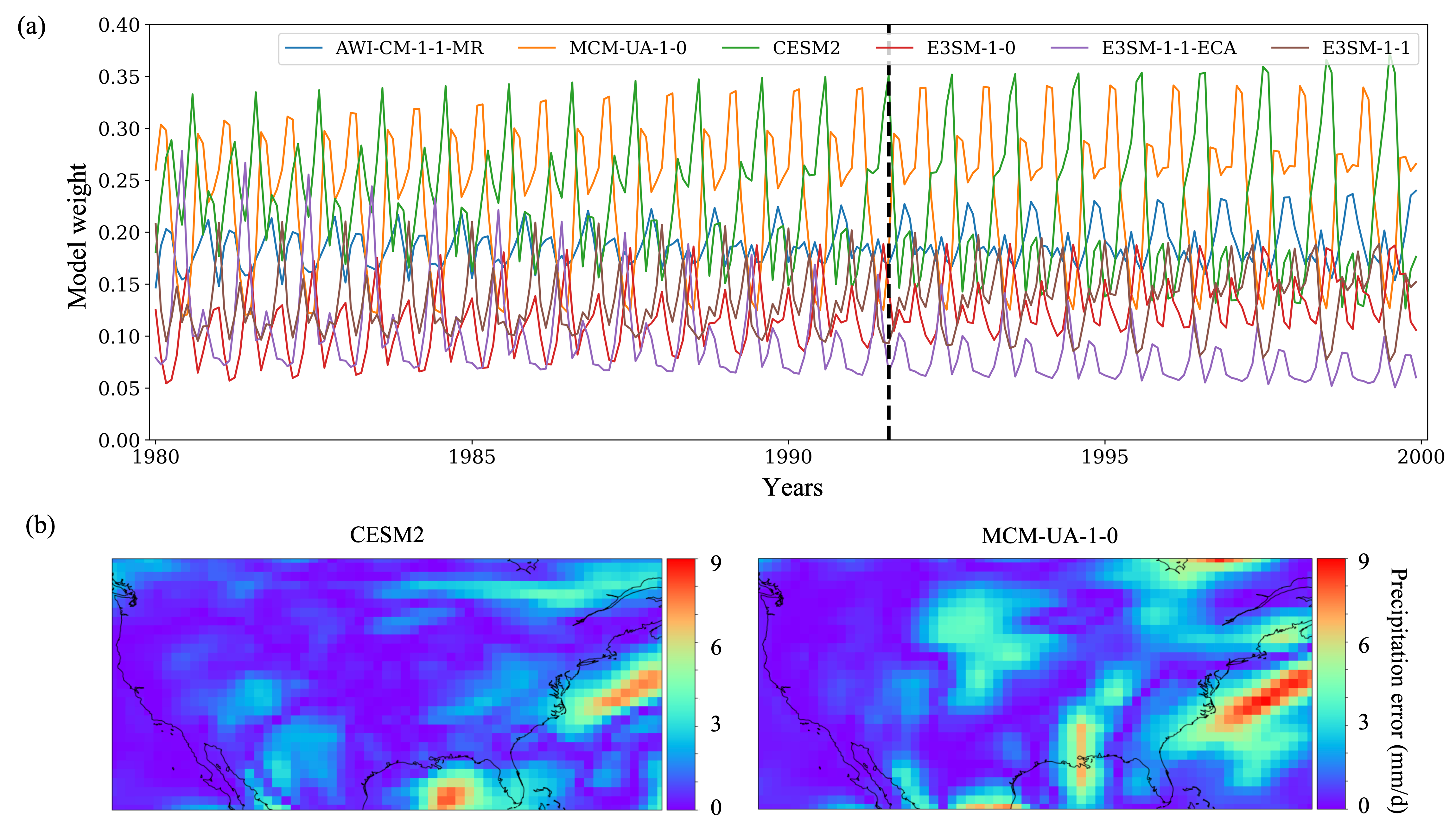}
\vspace{-0.4cm}
\caption{(a) Spatially averaged model weights over the simulation domain in the 20 year  period for the six GCMs in the synthetic study; (b) Prediction errors of model CESM2 and MCM-UA-1-0 in the timestamp of August 1991 as highlighted in the black line of (a)}
\label{fig:temporal}
\end{figure}

The superior prediction performance of our BNN is partially attributed to its spatiotemporally varying weights. Figure~\ref{fig:perform2}(b) presents the temporally averaged weights over the test period for the six individual GCMs in CONUS. We can see that the three top-performing models---AWI-CM-1-1-MR, MCM-UA-1-0, and CESM2---receive higher weights than the others overall, and the weights in each individual GCM vary spatially.       
We then divided the simulation domain into four regions (see Figure~\ref{fig:S2_data}(a)) to closely examine the BNN's spatial weighting and investigate whether its weighting aligns with the GCM's skill. Figure~\ref{fig:spatial}(a) summarizes the temporally averaged weights in the four regions and the entire CONUS domain for the six individual models, and it also presents the equal weights as a baseline.  
The figure indicates that although models MCM-UA-1-0 and CESM2 have the highest weights overall in CONUS, MCM-UA-1-0 contributes highly in the West, and CESM2 is the dominant GCM in the North and East. This spatially varying weight aligns well with each individual model's spatially varying skill. Take the West region, for example: Figure~\ref{fig:spatial}(b) indicates that model MCM-UA-1-0 performs better than E3SM-1-0 with smaller prediction errors in the West, and the BNN also assigns a higher weight to MCM-UA-1-0 in this region. This suggests that the BNN's spatially varying weighting reasonably reflects GCMs' geographically heterogeneous prediction skill.    
Additionally, we investigate the BNN's temporally varying weights. Figure~\ref{fig:temporal}(a) plots the spatially averaged weights of the 20 years for the six individual models. The figure indicates that all the models present a seasonally changing weight, and no individual model performs the best all the time. This suggests the importance of calculating temporally varying weights in the ensembling. We picked a timestamp, August 1991, for a detailed analysis and present the absolute prediction errors of model CESM2 and MCM-UA-1-0 at this specific time in Figure~\ref{fig:temporal}(b). The figure indicates that model CESM2 predicts more accurate precipitation in August 1991 than MCM-UA-1-0 by producing smaller prediction errors. Moreover, the BNN accurately estimates the temporal-aware weights by assigning a larger value to model CESM2 at this time step, which reasonably leverages the model's seasonally distinct skill.  

By calculating the spatiotemporally varying weights that accurately reflect the individual model's diverse skill across space and time, our BNN provides interpretability about which model contributes more to the ensemble prediction in which region and at which time. This scientific insight improves our understanding of each GCM's predictive performance and help the model development by leveraging each model's merits. For example, BNN identified that the model MCM-UA-1-0 is more accurate in predicting the precipitation in the West region of CONUS, and CESM2 performs better in the Summer season. Then we can go back to explore the mechanisms of the two models to investigate why they yield better performance in the specific region at the specific time. On the other hand, we can also examine why a certain model performs poorly in a certain region at a certain time. Combining this comprehensive analysis, we can take advantage of each individual model's strength to build a more powerful GCM for precipitation prediction. And we can also explain that the BNN results in the superior ensemble predictions because it assigns higher weights to the regions and times where the model performs better. In essence, we are confident in that we are getting right answers for the right reasons. 

\begin{figure}[h]
\centering
\includegraphics[scale=0.4]{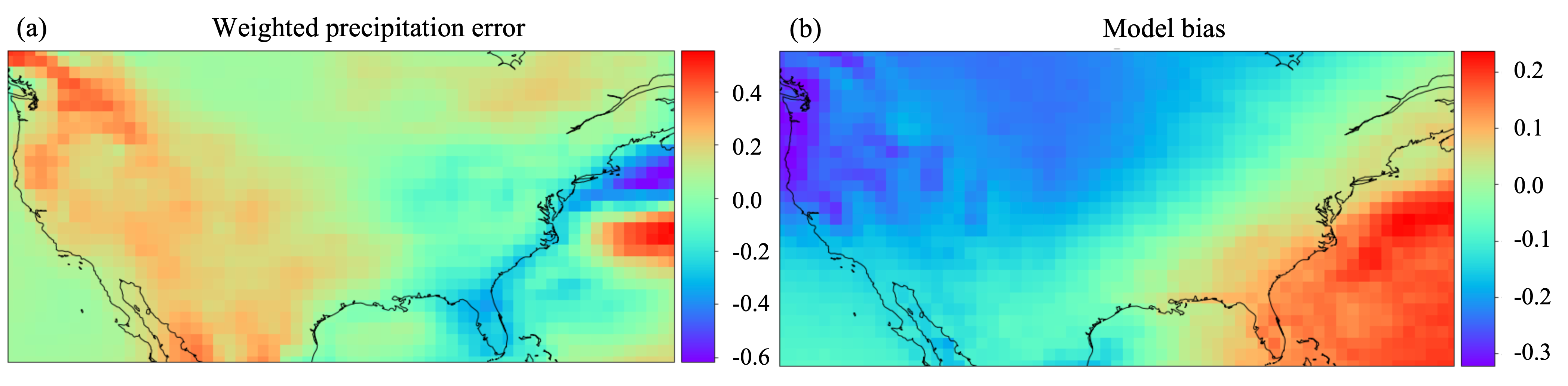}
\vspace{-0.2cm}
\caption{(a) Weighted precipitation errors (mm/d) of the six GCMs in the synthetic study; (b) The estimated bias (mm/d) ($\beta$ in Figure~\ref{fig:archi}) of BNN to compensate the weighted precipitation errors to enable a better ensemble prediction.}
\label{fig:bias}
\end{figure}

Besides the smart weighting scheme, the spatially varying bias term in the BNN ensembling also plays an important role for accurate precipitation prediction. As shown in Figure~\ref{fig:bias}(a), which presents the weighted prediction errors of the six GCMs, the northwest region has a relatively large positive prediction error. To compensate for the error and make the ensemble prediction fit the calibration data well, the BNN estimates the bias with a relatively large negative value in the region, as depicted in Figure~\ref{fig:bias}(b). This bias compensation scheme is particularly important when all the individual GCMs generate overestimation or underestimation, in which case the ensemble prediction will hardly perform better than the best-performing individual GCMs despite the ensembling schemes. In this situation, by introducing the bias term and calibrating its value against the data, we can improve the ensemble predictions. Additionally, this bias term is a function of space and time, so its calculation reflects the spatiotemporally varying model skill.

\begin{figure}[h]
\centering
\includegraphics[scale=0.4]{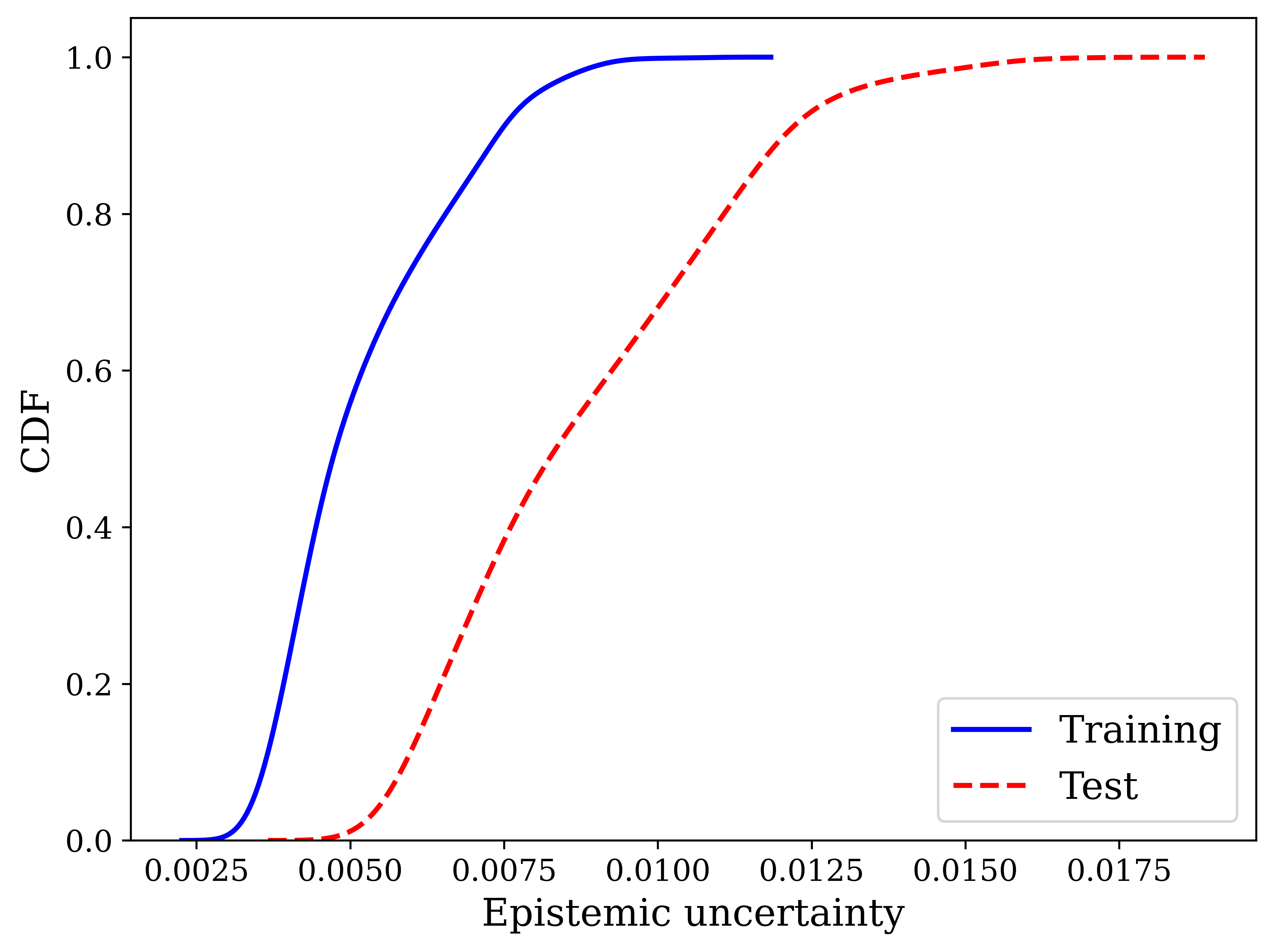}
\vspace{-0.2cm}
\caption{Epistemic uncertainty of the training and out-of-sample test data calculated by BNN in the synthetic study.}
\label{fig:noise}
\end{figure}

The BNN performs ensemble prediction in the Bayesian context, so it can quantify the data uncertainty to consider the data noise and quantify the epistemic uncertainty to consider the extrapolation error. Because the ``observations'' come from model simulation data in this synthetic study, we do not have data noise. We focus more on the epistemic uncertainty discussion. Figure~\ref{fig:noise} shows the cumulative density function (CDF) of the epistemic uncertainty for the training and out-of-sample test data. The figure indicates that the BNN can reasonably quantify the uncertainty, where the epistemic uncertainty of the test data in the extrapolation regime is greater than that of the training data. This is highly desirable behavior and crucial in practice to prevent overconfident projection in the future climate. 
 
In this synthetic study, we demonstrate that our BNN produces superior prediction performance compared to that of the three state-of-the-art ensembling methods. BNN can accurately calculate the spatiotemporally varying model weights and biases, which can be justified by the model's prediction skill. This spatiotemporal-aware weighting scheme meanwhile provides the interpretability of the BNN to help us understand which models contribute more to the ensemble prediction at which locations and times. Additionally, we demonstrate that the BNN can reasonably quantify the epistemic uncertainty by producing a larger uncertainty bound in the extrapolation regime to avoid overconfident predictions.

\subsection{A Real Case Application}\label{sec:real} 
\begin{figure}[h]
\centering
\includegraphics[width=1\textwidth]{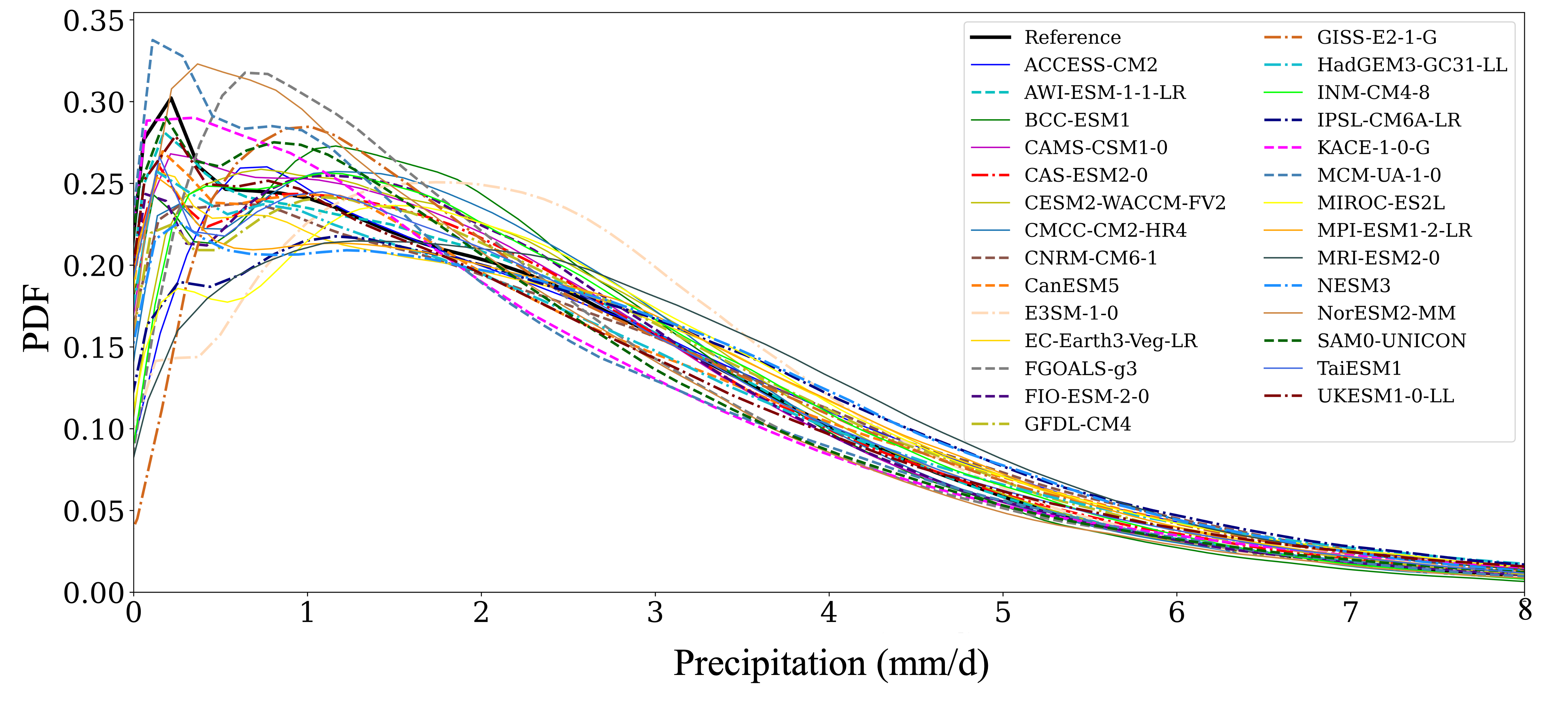}
\vspace{-0.5cm}
\caption{The PDFs of the precipitation data from the 28 GCMs for ensemble analysis and the reference data from the ERA5 reanalysis product.}
\label{fig:Precip_28models}
\end{figure}

After verification and validation of the BNN method, we applied it to a real case study for precipitation prediction where the ``observations'' come from the ERA5 reanalysis data. We considered the 53 GCMs from CMIP6 as the model set; these are described in Section~\ref{sec:data}.
The 53 GCMs are from 28 institutes. Given that the models from the same institute have strong dependence/similarities, we first performed data screening by selecting one model from one institute to roughly consider the model independence before the ensemble analysis \cite{leduc2016institutional,ashfaq2022evaluation}. For the models in the same institute, we chose the one with the smallest RMSE compared to the ERA5 reference data. The final selected 28 models are highlighted in bold in Table~\ref{tab:model}. Figure~\ref{fig:Precip_28models} shows the PDFs of the precipitation data for the 28 GCMs and the ERA5 reference. As shown, the 28 GCMs produce different precipitation simulations, and the major difference happens at the small precipitation values ($\le 2mm/d$). Some GCMs have close PDFs to the ``observations,'' and some others deviate significantly from the reference.   

We applied the BNN to the 28 GCMs for ensemble predictions and investigated whether our model can leverage each individual model's spatiotemporally varying skill to produce an accurate prediction---and meanwhile reasonably quantify the predictive uncertainty. We used the first 20 years of data for training and the remaining 15 years for out-of-sample testing. In the training, we used the ERA5 data for model calibration; in the testing, the ERA5 data were used as reference to evaluate the prediction performance. In the following results discussion, we first evaluate the BNN's prediction accuracy in comparison with the three baseline ensembling methods. Next, we analyze the BNN's model weights across the space and time and examine its interpretability. Lastly, we present the UQ results and discuss the computational costs.

Table~\ref{tab:rmse} summarizes the RMSEs of the 15 years' test data in the entire simulation domain and the four regions for the four ensembling methods, where the four regions are divided in the same manner as shown in Figure~\ref{fig:S2_data}(a). The BNN produces the smallest RMSEs in CONUS and in the West, North, and South regions, demonstrating the best prediction performance. Additionally, the BNN produced consistently smaller prediction errors than the simple average method, whereas in some cases, the advanced weighted average and spatially weighted average even produced larger RMSEs than the simple average. Please note that the ensembling results of the weighted average and spatially weighted average methods come from a fine-tuning of their hyperparameters, whereas for the BNN approach, we did not perform a hyperparameter and architecture optimization. A further improvement in the BNN prediction performance is expected after a better choice of its hyperparameters and network architectures.

\FloatBarrier
\begin{table*}[h]
\caption{The RMSEs of the 15 years' precipitation data (mm/d) in the test period at CONUS and the four sub-regions (Figure~\ref{fig:S2_data}(a)) for the four ensembling methods.}\label{tab:rmse}
\begin{tabular}{ p{2cm} p{3cm} p{3cm} p{3cm} p{1cm}}
\hline
  & Simple average  & Weighted average & Spatially weighted average & BNN\\
\hline
CONUS & 1.48 & 1.48 & 1.51 & 1.45\\
West  & 0.71 & 0.68 & 0.67 & 0.57\\
North & 0.25 & 0.27 & 0.28 & 0.23\\
South & 0.50 & 0.48 & 0.49 & 0.43\\
East  & 0.46 & 0.44 & 0.43 & 0.44\\
\hline
\end{tabular}

\end{table*}
\FloatBarrier

\begin{figure}[h]
\centering
\includegraphics[width=1\textwidth]{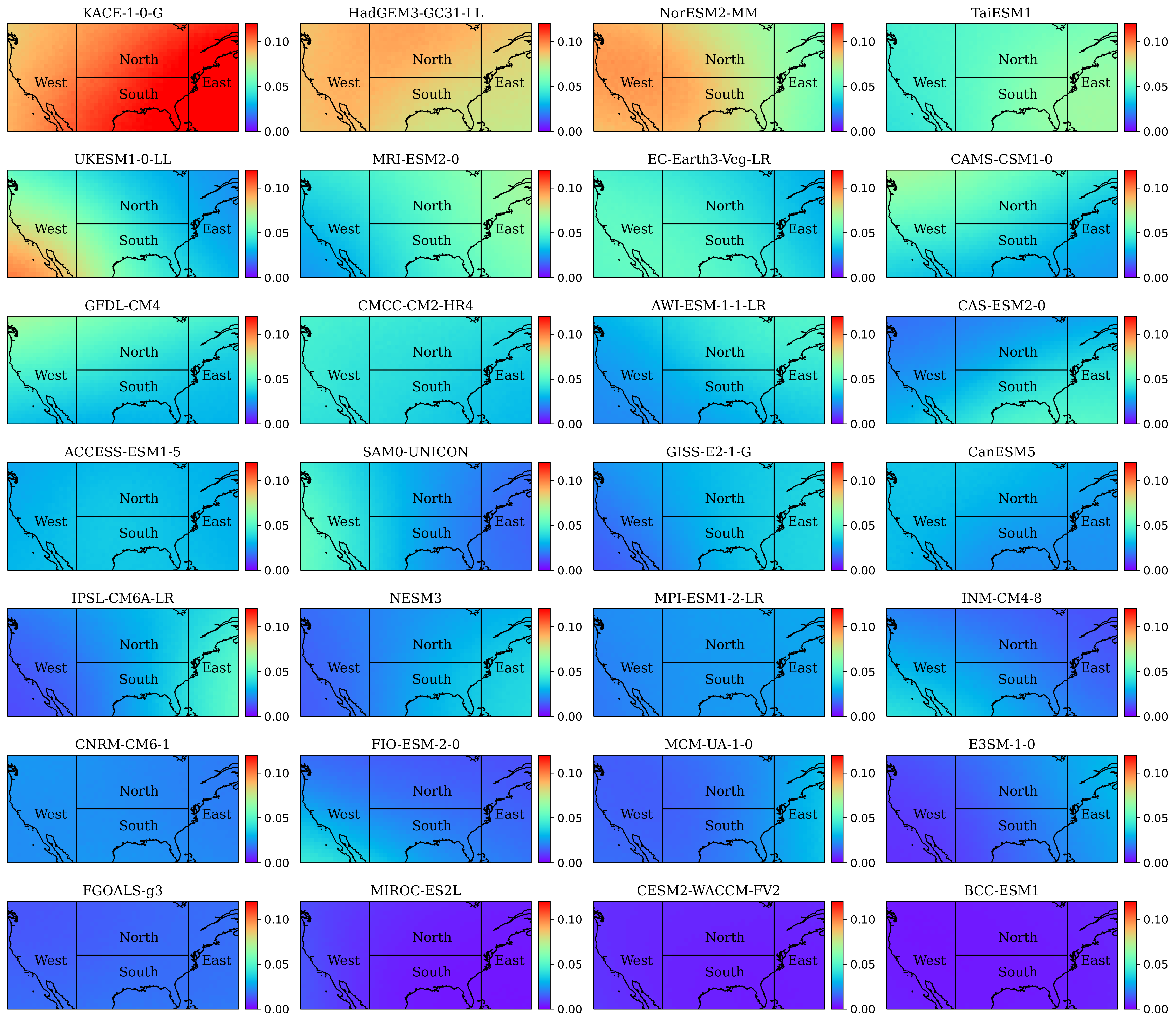}
\vspace{-0.4cm}
\caption{Temporally averaged model weights over the 15 years of the test period in CONUS for the 28 GCMs in the real case application.}
\label{fig:c328models}
\end{figure}

The superior prediction performance of the BNN benefits from its spatiotemporal-aware weighting scheme.
Figure~\ref{fig:c328models} shows the temporally averaged model weights in CONUS for the 28 GCMs. We organized the models from the largest weights to the smallest weights in row-wise order. Each GCM presents geographically heterogeneous weights. Overall, model KACE-1-0-G, HadGEM3-GC31-LL, and NorESM2-MM in the top row gain the highest weights, and model MIROC-ES2L, CESM2-WACCM-FV2, and BCC-ESM1 on the bottom row obtain the lowest weights. However, the model's overall higher weight does not necessarily show a uniform higher weight across the domain at each grid cell. For example, in the second column of Figure~\ref{fig:c328models}, although model SAM0-UNICON has a smaller weight than MRI-ESM2-0 in most areas, it shows a higher weight in the West region.   

\begin{figure}[!ht]
\centering
\includegraphics[width=1\textwidth]{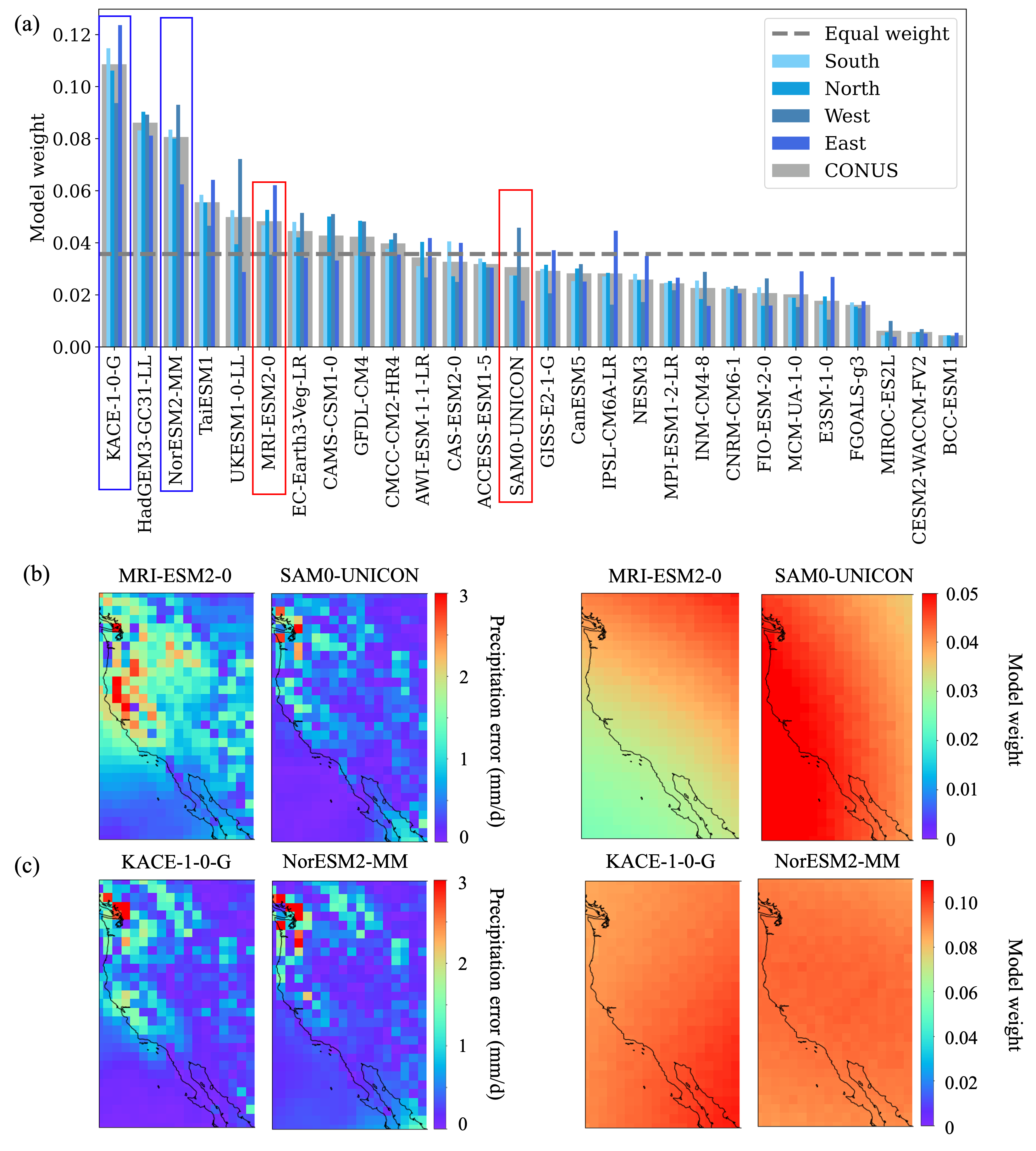}
\vspace{-0.4cm}
\caption{(a) Temporally averaged weights over the test period in CONUS and the four regions (see Figure \ref{fig:c328models}) for the 28 GCMs considered in the real case application; (b) Prediction errors and model weights of model MRI-ESM2-0 and SAM0-UNICON in the West region; (c) Prediction errors and model weights of model KACE-1-0-G and NorESM2-MM in the West region.}
\label{fig:c3spatial}
\end{figure}

Figure~\ref{fig:c3spatial}(a) summarizes the temporally averaged model weights in CONUS and the four regions for the 28 GCMs. We can clearly see that the models present regionally varying weights. Ten models have a higher weight than the equal value (i.e., $1/28$). For some models whose overall weights are below the equal weight, they could still show a higher weight in a certain region. 
For example, model SAM0-UNICON presents a larger weight in the West region than MRI-ESM2-0, although its overall weight in CONUS is smaller than the latter. This spatially varying weight assignment is consistent with the individual model's prediction skill. As shown in Figure \ref{fig:c3spatial}(b), model SAM0-UNICON shows a smaller prediction error than MRI-ESM2-0 in the West, where the RMSEs of SAM0-UNICON and MRI-ESM2-0 are 0.7 and 1.2, respectively;  we also observe a higher spatial weight of SAM0-UNICON in this region.
Additionally, Figure~\ref{fig:c3spatial}(c) illustrates that model KACE-1-0-G and NorESM2-MM have  similar weights in the West region, and  these two models indeed show similar prediction performance: the RMSE of KACE-1-0-G is 0.65 close to that of the NorESM2-MM  value of 0.63.

\begin{figure}[!ht]
\centering
\includegraphics[width=1\textwidth]{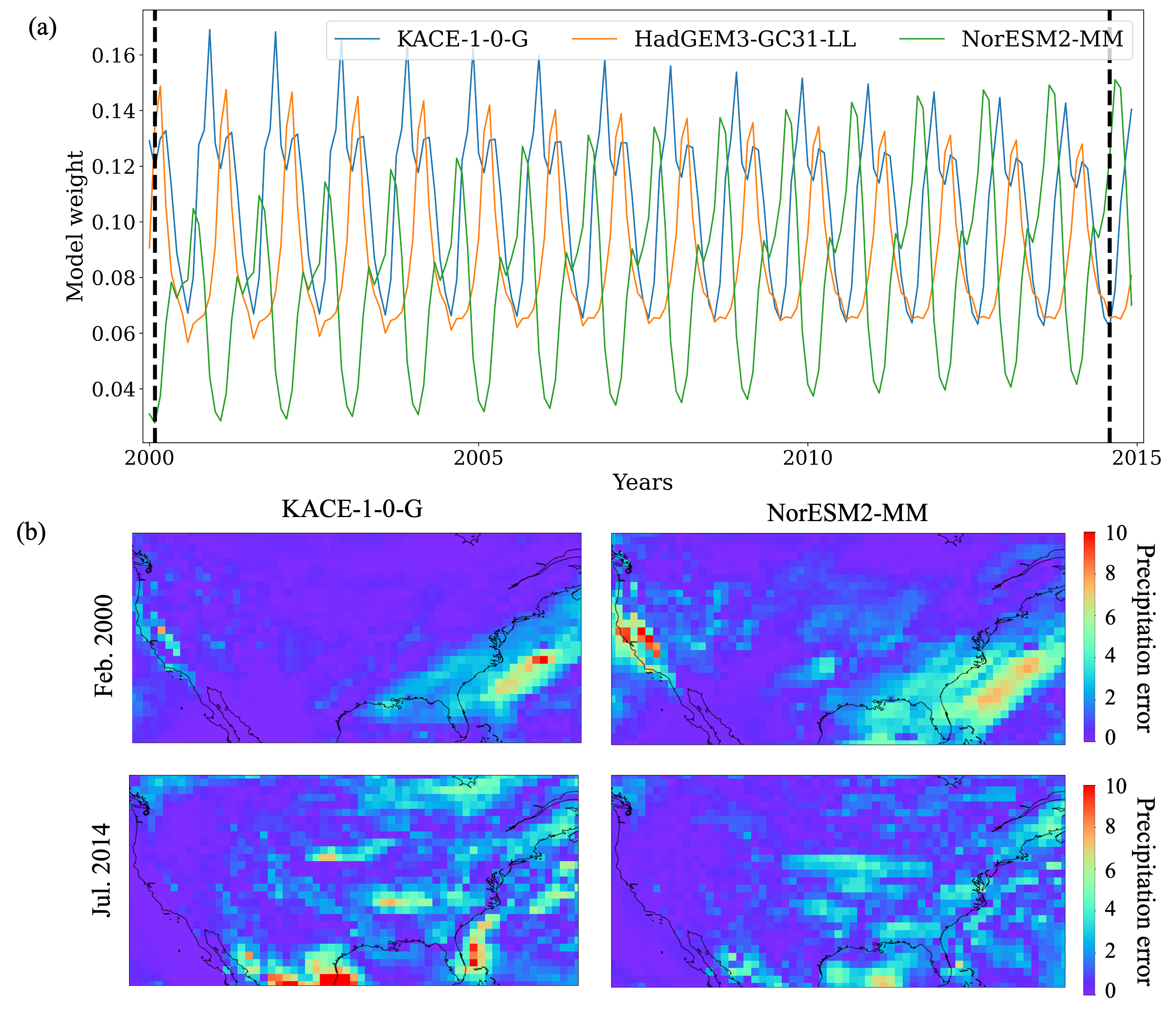}
\vspace{-0.2cm}
\caption{(a) Spatially averaged weights in the 15 years of test period for the three top performing GCMs in the real case application; (b) Prediction errors (mm/d) of model KACE-1-0-G and NorESM2-MM in February 2000 and July 2014.}
\label{fig:c3temporal}
\end{figure}

The BNN not only gives reasonable spatially aware weights that accurately reflect the individual model's spatially varying skill, but it also produces skill-consistent weights in the temporal dimension. To avoid a busy figure and for a better demonstration,  Figure~\ref{fig:c3temporal}(a) plots the spatially averaged model weights in the out-of-sample test period for three GCMs, which show the top prediction performance and have the highest spatial weights. All of the three models demonstrate a seasonally changing weight, and none of them obtain the highest weights all the time.
The weights of model KACE-1-0-G show a decreasing annual trend regardless of the seasonality, the weights of model NorESM2-MM present an increasing annual trend, and there is no much annual change in the weights of model HadGEM3-GC31-LL. We picked two timestamps---at the beginning and at the end of the test period---to analyze the weights of model KACE-1-0-G and NorESM2-MM in detail. Figure~\ref{fig:c3temporal}(b) shows that KACE-1-0-G has a smaller prediction error than NorESM2-MM in February 2000, which justifies its higher model weights at this time. Additionally, the lower weight of KACE-1-0-G in July 2014 once again aligns with its relatively higher prediction errors at this time. 

\begin{figure}[!ht]
\centering
\includegraphics[scale=0.4]{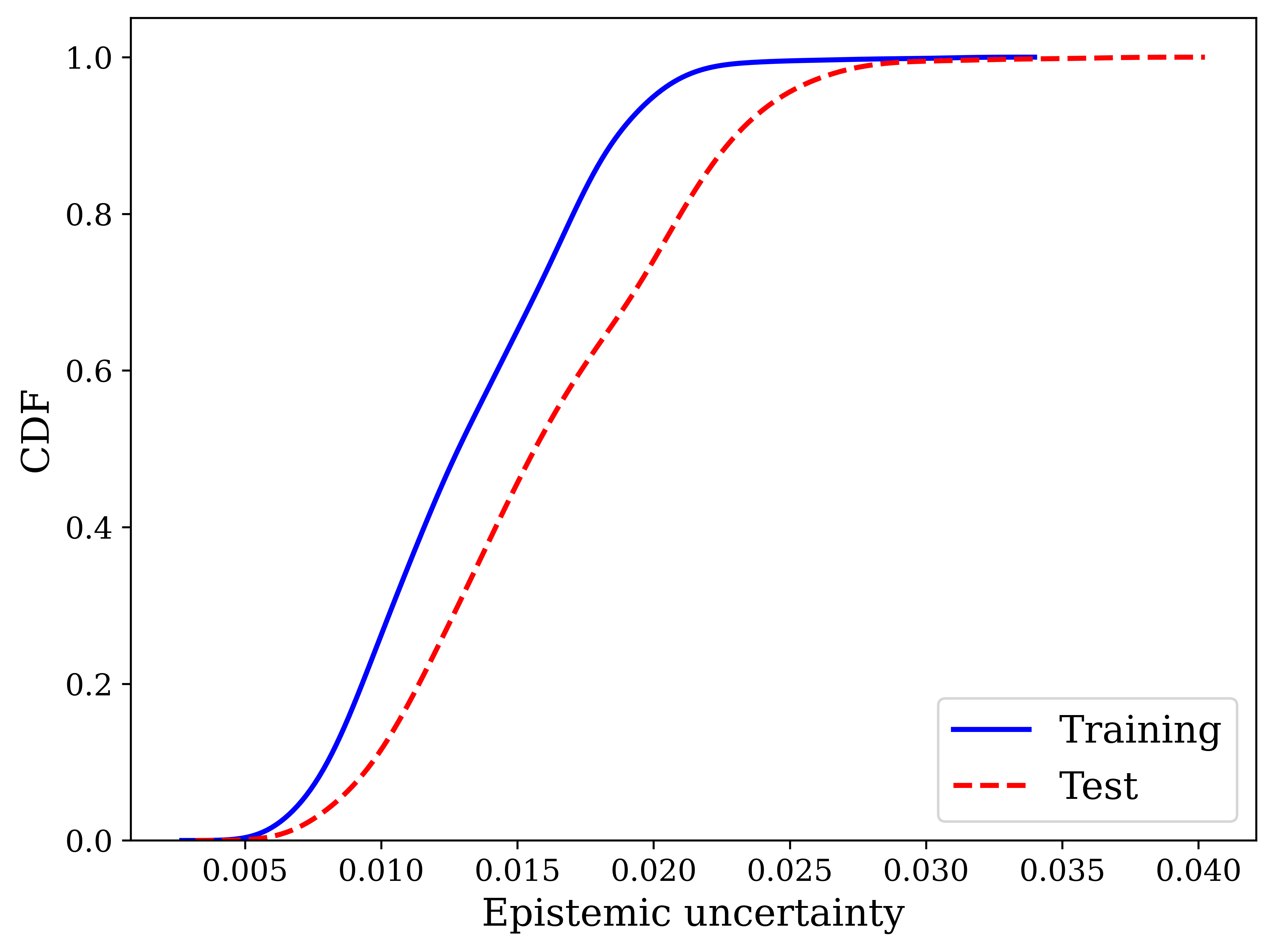}
\vspace{-0.2cm}
\caption{Epistemic uncertainty of the training and out-of-sample test data calculated by BNN in the real case application.}
\label{fig:uq3}
\end{figure}

The BNN accurately calculates the model weights for each individual model in each grid cell at each time step. Its weighting scheme sufficiently leverages models' prediction skill and produces skill-consistent weights. This smart weighting not only improves model prediction accuracy, but it also provides interpretability about each GCM's contribution to the ensemble prediction. Please note that all the results and weights analysis presented in this real case application are based on the out-of-sample test data, so when deploying the BNN method in practice for future projection where the ground truth is unknown, its verified interpretable and skill-consistent weights increase our confidence in the BNN's ensemble prediction.    
Certainly, when projecting to the future unknown conditions, besides the point estimate, we are also interested in the predictive uncertainty. BNN can reasonably quantify the epistemic uncertainty caused by the model ignorance and data shortage. Figure~\ref{fig:uq3} shows the CDFs of the epistemic uncertainty for the training and out-of-sample test data. The figure indicates that BNN produces a larger epistemic uncertainty of the test data than that of the training data, accurately reflecting our lesser confidence in the unknown conditions and thus preventing overconfident extrapolation.

In this real case application, we successfully apply the BNN ensembling method to 28 GCMs from CMIP6 for precipitation predictions in CONUS. We demonstrate BNN's superior prediction performance regionally and locally in comparison to the three baseline methods. We investigate BNN's spatiotemporal-aware weighting scheme, verify its weight's consistency with the model prediction skill, and interpret the individual models' contribution to the ensemble prediction spatially and temporally. Lastly, we analyze the reasonableness of BNN's UQ capability. 

One possible limitation of the BNN ensembling scheme is the high computational cost. In this work, all the training ends at 1000 epochs when the loss function shows marginal decay. In the synthetic study in which six GCMs are considered, it takes about 35 minutes to train one NN and 29.17 hours to finish the training of 50 NNs in the BNN ensembling. For the real case application where 28 GCMs are analyzed, it takes about 40 minutes to train one NN and 33.33 hours to train the 50 NNs. All the experiments were performed on a 2.3 GHz Quad-Core Intel Core i7 CPU. Roughly speaking, the computational cost increases with increasing numbers of  networks in BNN training and ensemble GCMs, as well as with the resolution of the GCMs; this is because the BNN calculates weights at each time step in each grid cell. 
In spite of the relatively high computational cost of the BNN compared to other ensembling schemes, the cost is affordable (e.g., within one or two days); more importantly, the BNN provides better prediction performance, interpretable ensembling results, and UQ.


\section{Conclusions and Future Work}
In this work, we propose a BNN ensembling method for multiple model analysis to enhance the predictive capability. The method improves prediction accuracy by learning spatiotemporally varying model weights and biases based on the individual models' skill in simulating the observations across space and time. Additionally, the BNN method accounts for the varying quality of the observations by incorporating their aleatoric uncertainty and avoids overconfident extrapolating predictions by quantifying the epistemic uncertainty. More importantly, the method offers interpretability about which models contribute more to the ensemble prediction at which locations and seasons. This insight advances predictive understanding, guides process-based model development, and prioritizes data collection.

We apply the BNN ensembling method for precipitation prediction in CONUS based on the GCMs from CMIP6. In both synthetic and real case studies, we demonstrate that the BNN produces a better prediction performance than the three baseline ensembling approaches; it can correctly assign a higher weight to the regions and the seasons where the individual GCM fits the ``observations'' better; and it gives a reasonable bias value to compensate for the error of the weighted average to enable a better ensemble prediction than the individual models.   
Additionally, we verify that the proposed BNN's interpretability is consistent with our prior knowledge in the synthetic design and with our understanding of localized GCM performance in the real case application. Finally, the BNN shows an increasing uncertainty when the prediction is farther away from the period with constrained data, which appropriately reflects our predictive confidence and the trustworthiness of the models in the changing climate. 
Although the BNN ensembling method produces high-quality, interpretable, and uncertainty-aware predictions at the expense of high computational costs in calculating the grid-specific and time-specific model weights and biases, the cost is affordable: for example, about 33 hours are spent in application of the 28 GCMs. More importantly, the provided high predictive accuracy and the insights of the model performance are significant.      
In the future, we will apply the BNN ensembling technique for other Earth system modeling problems, including predictions of other response variables from the GCMs and problems in other disciplines such as hydrology and ecology.

\section{Data Availability Statement}
The model simulation and reanalysis reference data can be freely downloaded from the publicly accessible website as indicated in Section~\ref{sec:data}. The BNN code used in this study can be found in the first author's personal GitHub, and the link will be made public after the paper is published.

\section{Author Contributions}
MF implemented the numerical experiments, prepared the figures and analyzed the results. DL developed the algorithms, contributed to the research plan, and interpreted the results. DR processed the data. EMP formulated the problem and interpreted the results. All the four authors contributed to the manuscript writing.

\acknowledgments
This research was supported by the Artificial Intelligence Initiative as part of the Laboratory Directed Research and Development Program of Oak Ridge National Laboratory, managed by UT-Battelle, LLC, for the US DOE under contract DE-AC05-00OR22725. It is also sponsored by the Data-Driven Decision Control for Complex Systems (DnC2S) project funded by the US DOE, Office of Advanced Scientific Computing Research and the Critical Interfaces Science Focus Area project funded by the US DOE, Office of Biological and Environmental Research. 

\newpage

\bibliography{Ref.bib}

\providecommand{\noopsort}[1]{}\providecommand{\singleletter}[1]{#1}%
\begin{thebibliography}{}

\bibitem [\protect \citeauthoryear {%
Abramowitz%
\ \BBA {} Bishop%
}{%
Abramowitz%
\ \BBA {} Bishop%
}{%
{\protect \APACyear {2015}}%
}]{%
abramowitz2015climate}
\APACinsertmetastar {%
abramowitz2015climate}%
\begin{APACrefauthors}%
Abramowitz, G.%
\BCBT {}\ \BBA {} Bishop, C.%
\end{APACrefauthors}%
\unskip\
\newblock
\APACrefYearMonthDay{2015}{}{}.
\newblock
{\BBOQ}\APACrefatitle {Climate model dependence and the ensemble dependence
  transformation of CMIP projections} {Climate model dependence and the
  ensemble dependence transformation of cmip projections}.{\BBCQ}
\newblock
\APACjournalVolNumPages{Journal of Climate}{28}{6}{2332--2348}.
\PrintBackRefs{\CurrentBib}

\bibitem [\protect \citeauthoryear {%
Ahmed%
\ \protect \BOthers {.}}{%
Ahmed%
\ \protect \BOthers {.}}{%
{\protect \APACyear {2020}}%
}]{%
Ahmed2020}
\APACinsertmetastar {%
Ahmed2020}%
\begin{APACrefauthors}%
Ahmed, K.%
, Sachindra, D.%
, Shahid, S.%
, Iqbal, Z.%
, Nawaz, N.%
\BCBL {}\ \BBA {} Khan, N.%
\end{APACrefauthors}%
\unskip\
\newblock
\APACrefYearMonthDay{2020}{}{}.
\newblock
{\BBOQ}\APACrefatitle {Multi-model ensemble predictions of precipitation and
  temperature using machine learning algorithms} {Multi-model ensemble
  predictions of precipitation and temperature using machine learning
  algorithms}.{\BBCQ}
\newblock
\APACjournalVolNumPages{Atmospheric Research}{236}{}{104806}.
\newblock
\begin{APACrefURL}
  \url{https://www.sciencedirect.com/science/article/pii/S0169809519309858}
  \end{APACrefURL}
\newblock
\begin{APACrefDOI} \doi{https://doi.org/10.1016/j.atmosres.2019.104806}
  \end{APACrefDOI}
\PrintBackRefs{\CurrentBib}

\bibitem [\protect \citeauthoryear {%
Alexander%
\ \BBA {} Easterbrook%
}{%
Alexander%
\ \BBA {} Easterbrook%
}{%
{\protect \APACyear {2015}}%
}]{%
alexander2015software}
\APACinsertmetastar {%
alexander2015software}%
\begin{APACrefauthors}%
Alexander, K.%
\BCBT {}\ \BBA {} Easterbrook, S\BPBI M.%
\end{APACrefauthors}%
\unskip\
\newblock
\APACrefYearMonthDay{2015}{}{}.
\newblock
{\BBOQ}\APACrefatitle {The software architecture of climate models: a graphical
  comparison of CMIP5 and EMICAR5 configurations} {The software architecture of
  climate models: a graphical comparison of cmip5 and emicar5
  configurations}.{\BBCQ}
\newblock
\APACjournalVolNumPages{Geoscientific Model Development}{8}{4}{1221--1232}.
\PrintBackRefs{\CurrentBib}

\bibitem [\protect \citeauthoryear {%
Amos%
\ \protect \BOthers {.}}{%
Amos%
\ \protect \BOthers {.}}{%
{\protect \APACyear {2020}}%
}]{%
amos2020projecting}
\APACinsertmetastar {%
amos2020projecting}%
\begin{APACrefauthors}%
Amos, M.%
, Young, P\BPBI J.%
, Hosking, J\BPBI S.%
, Lamarque, J\BHBI F.%
, Abraham, N\BPBI L.%
, Akiyoshi, H.%
\BDBL {}others%
\end{APACrefauthors}%
\unskip\
\newblock
\APACrefYearMonthDay{2020}{}{}.
\newblock
{\BBOQ}\APACrefatitle {Projecting ozone hole recovery using an ensemble of
  chemistry--climate models weighted by model performance and independence}
  {Projecting ozone hole recovery using an ensemble of chemistry--climate
  models weighted by model performance and independence}.{\BBCQ}
\newblock
\APACjournalVolNumPages{Atmospheric Chemistry and Physics}{20}{16}{9961--9977}.
\PrintBackRefs{\CurrentBib}

\bibitem [\protect \citeauthoryear {%
Ashfaq%
, Rastogi%
, Abid%
\BCBL {}\ \BBA {} Kao%
}{%
Ashfaq%
\ \protect \BOthers {.}}{%
{\protect \APACyear {2022}}%
}]{%
ashfaq2022evaluation}
\APACinsertmetastar {%
ashfaq2022evaluation}%
\begin{APACrefauthors}%
Ashfaq, M.%
, Rastogi, D.%
, Abid, M\BPBI A.%
\BCBL {}\ \BBA {} Kao, S\BHBI C.%
\end{APACrefauthors}%
\unskip\
\newblock
\APACrefYearMonthDay{2022}{}{}.
\newblock
{\BBOQ}\APACrefatitle {Evaluation of CMIP6 GCMs over the CONUS for downscaling
  studies} {Evaluation of cmip6 gcms over the conus for downscaling
  studies}.{\BBCQ}
\newblock

\PrintBackRefs{\CurrentBib}

\bibitem [\protect \citeauthoryear {%
Bishop%
\ \BBA {} Abramowitz%
}{%
Bishop%
\ \BBA {} Abramowitz%
}{%
{\protect \APACyear {2013}}%
}]{%
bishop2013climate}
\APACinsertmetastar {%
bishop2013climate}%
\begin{APACrefauthors}%
Bishop, C\BPBI H.%
\BCBT {}\ \BBA {} Abramowitz, G.%
\end{APACrefauthors}%
\unskip\
\newblock
\APACrefYearMonthDay{2013}{}{}.
\newblock
{\BBOQ}\APACrefatitle {Climate model dependence and the replicate Earth
  paradigm} {Climate model dependence and the replicate earth paradigm}.{\BBCQ}
\newblock
\APACjournalVolNumPages{Climate dynamics}{41}{3}{885--900}.
\PrintBackRefs{\CurrentBib}

\bibitem [\protect \citeauthoryear {%
Brunner%
, Lorenz%
, Zumwald%
\BCBL {}\ \BBA {} Knutti%
}{%
Brunner%
\ \protect \BOthers {.}}{%
{\protect \APACyear {2019}}%
}]{%
brunner2019quantifying}
\APACinsertmetastar {%
brunner2019quantifying}%
\begin{APACrefauthors}%
Brunner, L.%
, Lorenz, R.%
, Zumwald, M.%
\BCBL {}\ \BBA {} Knutti, R.%
\end{APACrefauthors}%
\unskip\
\newblock
\APACrefYearMonthDay{2019}{}{}.
\newblock
{\BBOQ}\APACrefatitle {Quantifying uncertainty in European climate projections
  using combined performance-independence weighting} {Quantifying uncertainty
  in european climate projections using combined performance-independence
  weighting}.{\BBCQ}
\newblock
\APACjournalVolNumPages{Environmental Research Letters}{14}{12}{124010}.
\PrintBackRefs{\CurrentBib}

\bibitem [\protect \citeauthoryear {%
Dawei~Li%
\ \BBA {} Chen%
}{%
Dawei~Li%
\ \BBA {} Chen%
}{%
{\protect \APACyear {2021}}%
}]{%
Li2021}
\APACinsertmetastar {%
Li2021}%
\begin{APACrefauthors}%
Dawei~Li, Y\BPBI L.%
\BCBT {}\ \BBA {} Chen, C.%
\end{APACrefauthors}%
\unskip\
\newblock
\APACrefYearMonthDay{2021}{}{}.
\newblock
{\BBOQ}\APACrefatitle {MSDM v1. 0: A machine learning model for precipitation
  nowcasting over eastern China using multisource data} {Msdm v1. 0: A machine
  learning model for precipitation nowcasting over eastern china using
  multisource data}.{\BBCQ}
\newblock
\APACjournalVolNumPages{Geoscientific Model Development}{14}{6}{4019-4034}.
\PrintBackRefs{\CurrentBib}

\bibitem [\protect \citeauthoryear {%
Demory%
\ \protect \BOthers {.}}{%
Demory%
\ \protect \BOthers {.}}{%
{\protect \APACyear {2020}}%
}]{%
demory2020european}
\APACinsertmetastar {%
demory2020european}%
\begin{APACrefauthors}%
Demory, M\BHBI E.%
, Berthou, S.%
, Fern{\'a}ndez, J.%
, S{\o}rland, S\BPBI L.%
, Brogli, R.%
, Roberts, M\BPBI J.%
\BDBL {}others%
\end{APACrefauthors}%
\unskip\
\newblock
\APACrefYearMonthDay{2020}{}{}.
\newblock
{\BBOQ}\APACrefatitle {European daily precipitation according to EURO-CORDEX
  regional climate models (RCMs) and high-resolution global climate models
  (GCMs) from the High-Resolution Model Intercomparison Project (HighResMIP)}
  {European daily precipitation according to euro-cordex regional climate
  models (rcms) and high-resolution global climate models (gcms) from the
  high-resolution model intercomparison project (highresmip)}.{\BBCQ}
\newblock
\APACjournalVolNumPages{Geoscientific Model Development}{13}{11}{5485--5506}.
\PrintBackRefs{\CurrentBib}

\bibitem [\protect \citeauthoryear {%
Dinu Maria~Jose%
\ \BBA {} Dwarakish%
}{%
Dinu Maria~Jose%
\ \BBA {} Dwarakish%
}{%
{\protect \APACyear {2022}}%
}]{%
Jose2022}
\APACinsertmetastar {%
Jose2022}%
\begin{APACrefauthors}%
Dinu Maria~Jose, A\BPBI M\BPBI V.%
\BCBT {}\ \BBA {} Dwarakish, G\BPBI S.%
\end{APACrefauthors}%
\unskip\
\newblock
\APACrefYearMonthDay{2022}{}{}.
\newblock
{\BBOQ}\APACrefatitle {Improving multiple model ensemble predictions of daily
  precipitation and temperature through machine learning techniques} {Improving
  multiple model ensemble predictions of daily precipitation and temperature
  through machine learning techniques}.{\BBCQ}
\newblock
\APACjournalVolNumPages{Scientific Reports}{12}{1}{1-25}.
\PrintBackRefs{\CurrentBib}

\bibitem [\protect \citeauthoryear {%
Eyring%
\ \protect \BOthers {.}}{%
Eyring%
\ \protect \BOthers {.}}{%
{\protect \APACyear {2016}}%
}]{%
eyring2016overview}
\APACinsertmetastar {%
eyring2016overview}%
\begin{APACrefauthors}%
Eyring, V.%
, Bony, S.%
, Meehl, G\BPBI A.%
, Senior, C\BPBI A.%
, Stevens, B.%
, Stouffer, R\BPBI J.%
\BCBL {}\ \BBA {} Taylor, K\BPBI E.%
\end{APACrefauthors}%
\unskip\
\newblock
\APACrefYearMonthDay{2016}{}{}.
\newblock
{\BBOQ}\APACrefatitle {Overview of the Coupled Model Intercomparison Project
  Phase 6 (CMIP6) experimental design and organization} {Overview of the
  coupled model intercomparison project phase 6 (cmip6) experimental design and
  organization}.{\BBCQ}
\newblock
\APACjournalVolNumPages{Geoscientific Model Development}{9}{5}{1937--1958}.
\PrintBackRefs{\CurrentBib}

\bibitem [\protect \citeauthoryear {%
Eyring%
\ \protect \BOthers {.}}{%
Eyring%
\ \protect \BOthers {.}}{%
{\protect \APACyear {2019}}%
}]{%
eyring2019taking}
\APACinsertmetastar {%
eyring2019taking}%
\begin{APACrefauthors}%
Eyring, V.%
, Cox, P\BPBI M.%
, Flato, G\BPBI M.%
, Gleckler, P\BPBI J.%
, Abramowitz, G.%
, Caldwell, P.%
\BDBL {}others%
\end{APACrefauthors}%
\unskip\
\newblock
\APACrefYearMonthDay{2019}{}{}.
\newblock
{\BBOQ}\APACrefatitle {Taking climate model evaluation to the next level}
  {Taking climate model evaluation to the next level}.{\BBCQ}
\newblock
\APACjournalVolNumPages{Nature Climate Change}{9}{2}{102--110}.
\PrintBackRefs{\CurrentBib}

\bibitem [\protect \citeauthoryear {%
Fotheringham%
, Crespo%
\BCBL {}\ \BBA {} Yao%
}{%
Fotheringham%
\ \protect \BOthers {.}}{%
{\protect \APACyear {2015}}%
}]{%
fotheringham2015geographical}
\APACinsertmetastar {%
fotheringham2015geographical}%
\begin{APACrefauthors}%
Fotheringham, A\BPBI S.%
, Crespo, R.%
\BCBL {}\ \BBA {} Yao, J.%
\end{APACrefauthors}%
\unskip\
\newblock
\APACrefYearMonthDay{2015}{}{}.
\newblock
{\BBOQ}\APACrefatitle {Geographical and temporal weighted regression (GTWR)}
  {Geographical and temporal weighted regression (gtwr)}.{\BBCQ}
\newblock
\APACjournalVolNumPages{Geographical Analysis}{47}{4}{431--452}.
\PrintBackRefs{\CurrentBib}

\bibitem [\protect \citeauthoryear {%
Gleckler%
, Taylor%
\BCBL {}\ \BBA {} Doutriaux%
}{%
Gleckler%
\ \protect \BOthers {.}}{%
{\protect \APACyear {2008}}%
}]{%
gleckler2008performance}
\APACinsertmetastar {%
gleckler2008performance}%
\begin{APACrefauthors}%
Gleckler, P\BPBI J.%
, Taylor, K\BPBI E.%
\BCBL {}\ \BBA {} Doutriaux, C.%
\end{APACrefauthors}%
\unskip\
\newblock
\APACrefYearMonthDay{2008}{}{}.
\newblock
{\BBOQ}\APACrefatitle {Performance metrics for climate models} {Performance
  metrics for climate models}.{\BBCQ}
\newblock
\APACjournalVolNumPages{Journal of Geophysical Research:
  Atmospheres}{113}{D6}{}.
\PrintBackRefs{\CurrentBib}

\bibitem [\protect \citeauthoryear {%
Greve%
\ \protect \BOthers {.}}{%
Greve%
\ \protect \BOthers {.}}{%
{\protect \APACyear {2014}}%
}]{%
greve2014global}
\APACinsertmetastar {%
greve2014global}%
\begin{APACrefauthors}%
Greve, P.%
, Orlowsky, B.%
, Mueller, B.%
, Sheffield, J.%
, Reichstein, M.%
\BCBL {}\ \BBA {} Seneviratne, S\BPBI I.%
\end{APACrefauthors}%
\unskip\
\newblock
\APACrefYearMonthDay{2014}{}{}.
\newblock
{\BBOQ}\APACrefatitle {Global assessment of trends in wetting and drying over
  land} {Global assessment of trends in wetting and drying over land}.{\BBCQ}
\newblock
\APACjournalVolNumPages{Nature geoscience}{7}{10}{716--721}.
\PrintBackRefs{\CurrentBib}

\bibitem [\protect \citeauthoryear {%
Heinze-Deml%
, Sippel%
, Pendergrass%
, Lehner%
\BCBL {}\ \BBA {} Meinshausen%
}{%
Heinze-Deml%
\ \protect \BOthers {.}}{%
{\protect \APACyear {2021}}%
}]{%
Heinze2021}
\APACinsertmetastar {%
Heinze2021}%
\begin{APACrefauthors}%
Heinze-Deml, C.%
, Sippel, S.%
, Pendergrass, A\BPBI G.%
, Lehner, F.%
\BCBL {}\ \BBA {} Meinshausen, N.%
\end{APACrefauthors}%
\unskip\
\newblock
\APACrefYearMonthDay{2021}{}{}.
\newblock
{\BBOQ}\APACrefatitle {Latent Linear Adjustment Autoencoder v1.0: a novel
  method for estimating and emulating dynamic precipitation at high resolution}
  {Latent linear adjustment autoencoder v1.0: a novel method for estimating and
  emulating dynamic precipitation at high resolution}.{\BBCQ}
\newblock
\APACjournalVolNumPages{Geoscientific Model Development}{14}{8}{4977--4999}.
\newblock
\begin{APACrefURL} \url{https://gmd.copernicus.org/articles/14/4977/2021/}
  \end{APACrefURL}
\newblock
\begin{APACrefDOI} \doi{10.5194/gmd-14-4977-2021} \end{APACrefDOI}
\PrintBackRefs{\CurrentBib}

\bibitem [\protect \citeauthoryear {%
Karpechko%
, Maraun%
\BCBL {}\ \BBA {} Eyring%
}{%
Karpechko%
\ \protect \BOthers {.}}{%
{\protect \APACyear {2013}}%
}]{%
karpechko2013improving}
\APACinsertmetastar {%
karpechko2013improving}%
\begin{APACrefauthors}%
Karpechko, A\BPBI Y.%
, Maraun, D.%
\BCBL {}\ \BBA {} Eyring, V.%
\end{APACrefauthors}%
\unskip\
\newblock
\APACrefYearMonthDay{2013}{}{}.
\newblock
{\BBOQ}\APACrefatitle {Improving Antarctic total ozone projections by a
  process-oriented multiple diagnostic ensemble regression} {Improving
  antarctic total ozone projections by a process-oriented multiple diagnostic
  ensemble regression}.{\BBCQ}
\newblock
\APACjournalVolNumPages{Journal of the Atmospheric
  Sciences}{70}{12}{3959--3976}.
\PrintBackRefs{\CurrentBib}

\bibitem [\protect \citeauthoryear {%
Knutti%
, Furrer%
, Tebaldi%
, Cermak%
\BCBL {}\ \BBA {} Meehl%
}{%
Knutti%
\ \protect \BOthers {.}}{%
{\protect \APACyear {2010}}%
}]{%
knutti2010challenges}
\APACinsertmetastar {%
knutti2010challenges}%
\begin{APACrefauthors}%
Knutti, R.%
, Furrer, R.%
, Tebaldi, C.%
, Cermak, J.%
\BCBL {}\ \BBA {} Meehl, G\BPBI A.%
\end{APACrefauthors}%
\unskip\
\newblock
\APACrefYearMonthDay{2010}{}{}.
\newblock
{\BBOQ}\APACrefatitle {Challenges in combining projections from multiple
  climate models} {Challenges in combining projections from multiple climate
  models}.{\BBCQ}
\newblock
\APACjournalVolNumPages{Journal of Climate}{23}{10}{2739--2758}.
\PrintBackRefs{\CurrentBib}

\bibitem [\protect \citeauthoryear {%
Knutti%
\ \protect \BOthers {.}}{%
Knutti%
\ \protect \BOthers {.}}{%
{\protect \APACyear {2017}}%
}]{%
knutti2017climate}
\APACinsertmetastar {%
knutti2017climate}%
\begin{APACrefauthors}%
Knutti, R.%
, Sedl{\'a}{\v{c}}ek, J.%
, Sanderson, B\BPBI M.%
, Lorenz, R.%
, Fischer, E\BPBI M.%
\BCBL {}\ \BBA {} Eyring, V.%
\end{APACrefauthors}%
\unskip\
\newblock
\APACrefYearMonthDay{2017}{}{}.
\newblock
{\BBOQ}\APACrefatitle {A climate model projection weighting scheme accounting
  for performance and interdependence} {A climate model projection weighting
  scheme accounting for performance and interdependence}.{\BBCQ}
\newblock
\APACjournalVolNumPages{Geophysical Research Letters}{44}{4}{1909--1918}.
\PrintBackRefs{\CurrentBib}

\bibitem [\protect \citeauthoryear {%
Konapala%
, Mishra%
, Wada%
\BCBL {}\ \BBA {} Mann%
}{%
Konapala%
\ \protect \BOthers {.}}{%
{\protect \APACyear {2020}}%
}]{%
osti_1761643}
\APACinsertmetastar {%
osti_1761643}%
\begin{APACrefauthors}%
Konapala, G.%
, Mishra, A\BPBI K.%
, Wada, Y.%
\BCBL {}\ \BBA {} Mann, M\BPBI E.%
\end{APACrefauthors}%
\unskip\
\newblock
\APACrefYearMonthDay{2020}{6}{}.
\newblock
{\BBOQ}\APACrefatitle {Climate change will affect global water availability
  through compounding changes in seasonal precipitation and evaporation}
  {Climate change will affect global water availability through compounding
  changes in seasonal precipitation and evaporation}.{\BBCQ}
\newblock
\APACjournalVolNumPages{Nature Communications}{11}{1}{}.
\newblock
\begin{APACrefDOI} \doi{10.1038/s41467-020-16757-w} \end{APACrefDOI}
\PrintBackRefs{\CurrentBib}

\bibitem [\protect \citeauthoryear {%
Kumar%
, Kodra%
\BCBL {}\ \BBA {} Ganguly%
}{%
Kumar%
\ \protect \BOthers {.}}{%
{\protect \APACyear {2014}}%
}]{%
kumar2014regional}
\APACinsertmetastar {%
kumar2014regional}%
\begin{APACrefauthors}%
Kumar, D.%
, Kodra, E.%
\BCBL {}\ \BBA {} Ganguly, A\BPBI R.%
\end{APACrefauthors}%
\unskip\
\newblock
\APACrefYearMonthDay{2014}{}{}.
\newblock
{\BBOQ}\APACrefatitle {Regional and seasonal intercomparison of CMIP3 and CMIP5
  climate model ensembles for temperature and precipitation} {Regional and
  seasonal intercomparison of cmip3 and cmip5 climate model ensembles for
  temperature and precipitation}.{\BBCQ}
\newblock
\APACjournalVolNumPages{Climate dynamics}{43}{9}{2491--2518}.
\PrintBackRefs{\CurrentBib}

\bibitem [\protect \citeauthoryear {%
Leduc%
, Laprise%
, De~Elia%
\BCBL {}\ \BBA {} {\v{S}}eparovi{\'c}%
}{%
Leduc%
\ \protect \BOthers {.}}{%
{\protect \APACyear {2016}}%
}]{%
leduc2016institutional}
\APACinsertmetastar {%
leduc2016institutional}%
\begin{APACrefauthors}%
Leduc, M.%
, Laprise, R.%
, De~Elia, R.%
\BCBL {}\ \BBA {} {\v{S}}eparovi{\'c}, L.%
\end{APACrefauthors}%
\unskip\
\newblock
\APACrefYearMonthDay{2016}{}{}.
\newblock
{\BBOQ}\APACrefatitle {Is institutional democracy a good proxy for model
  independence?} {Is institutional democracy a good proxy for model
  independence?}{\BBCQ}
\newblock
\APACjournalVolNumPages{Journal of Climate}{29}{23}{8301--8316}.
\PrintBackRefs{\CurrentBib}

\bibitem [\protect \citeauthoryear {%
Lorenz%
\ \protect \BOthers {.}}{%
Lorenz%
\ \protect \BOthers {.}}{%
{\protect \APACyear {2018}}%
}]{%
lorenz2018prospects}
\APACinsertmetastar {%
lorenz2018prospects}%
\begin{APACrefauthors}%
Lorenz, R.%
, Herger, N.%
, Sedl{\'a}{\v{c}}ek, J.%
, Eyring, V.%
, Fischer, E\BPBI M.%
\BCBL {}\ \BBA {} Knutti, R.%
\end{APACrefauthors}%
\unskip\
\newblock
\APACrefYearMonthDay{2018}{}{}.
\newblock
{\BBOQ}\APACrefatitle {Prospects and caveats of weighting climate models for
  summer maximum temperature projections over North America} {Prospects and
  caveats of weighting climate models for summer maximum temperature
  projections over north america}.{\BBCQ}
\newblock
\APACjournalVolNumPages{Journal of Geophysical Research:
  Atmospheres}{123}{9}{4509--4526}.
\PrintBackRefs{\CurrentBib}

\bibitem [\protect \citeauthoryear {%
E.~Martin%
}{%
E.~Martin%
}{%
{\protect \APACyear {2018}}%
}]{%
martin2018future}
\APACinsertmetastar {%
martin2018future}%
\begin{APACrefauthors}%
Martin, E.%
\end{APACrefauthors}%
\unskip\
\newblock
\APACrefYearMonthDay{2018}{}{}.
\newblock
{\BBOQ}\APACrefatitle {Future projections of global pluvial and drought event
  characteristics} {Future projections of global pluvial and drought event
  characteristics}.{\BBCQ}
\newblock
\APACjournalVolNumPages{Geophysical Research Letters}{45}{21}{11--913}.
\PrintBackRefs{\CurrentBib}

\bibitem [\protect \citeauthoryear {%
G\BPBI M.~Martin%
, Klingaman%
\BCBL {}\ \BBA {} Moise%
}{%
G\BPBI M.~Martin%
\ \protect \BOthers {.}}{%
{\protect \APACyear {2017}}%
}]{%
martin2017connecting}
\APACinsertmetastar {%
martin2017connecting}%
\begin{APACrefauthors}%
Martin, G\BPBI M.%
, Klingaman, N\BPBI P.%
\BCBL {}\ \BBA {} Moise, A\BPBI F.%
\end{APACrefauthors}%
\unskip\
\newblock
\APACrefYearMonthDay{2017}{}{}.
\newblock
{\BBOQ}\APACrefatitle {Connecting spatial and temporal scales of tropical
  precipitation in observations and the MetUM-GA6} {Connecting spatial and
  temporal scales of tropical precipitation in observations and the
  metum-ga6}.{\BBCQ}
\newblock
\APACjournalVolNumPages{Geoscientific Model Development}{10}{1}{105--126}.
\PrintBackRefs{\CurrentBib}

\bibitem [\protect \citeauthoryear {%
Mueller%
\ \BBA {} Seneviratne%
}{%
Mueller%
\ \BBA {} Seneviratne%
}{%
{\protect \APACyear {2014}}%
}]{%
mueller2014systematic}
\APACinsertmetastar {%
mueller2014systematic}%
\begin{APACrefauthors}%
Mueller, B.%
\BCBT {}\ \BBA {} Seneviratne, S\BPBI I.%
\end{APACrefauthors}%
\unskip\
\newblock
\APACrefYearMonthDay{2014}{}{}.
\newblock
{\BBOQ}\APACrefatitle {Systematic land climate and evapotranspiration biases in
  CMIP5 simulations} {Systematic land climate and evapotranspiration biases in
  cmip5 simulations}.{\BBCQ}
\newblock
\APACjournalVolNumPages{Geophysical research letters}{41}{1}{128--134}.
\PrintBackRefs{\CurrentBib}

\bibitem [\protect \citeauthoryear {%
Mu{\~n}oz-Sabater%
\ \protect \BOthers {.}}{%
Mu{\~n}oz-Sabater%
\ \protect \BOthers {.}}{%
{\protect \APACyear {2021}}%
}]{%
munoz2021era5}
\APACinsertmetastar {%
munoz2021era5}%
\begin{APACrefauthors}%
Mu{\~n}oz-Sabater, J.%
, Dutra, E.%
, Agust{\'\i}-Panareda, A.%
, Albergel, C.%
, Arduini, G.%
, Balsamo, G.%
\BDBL {}others%
\end{APACrefauthors}%
\unskip\
\newblock
\APACrefYearMonthDay{2021}{}{}.
\newblock
{\BBOQ}\APACrefatitle {ERA5-Land: A state-of-the-art global reanalysis dataset
  for land applications} {Era5-land: A state-of-the-art global reanalysis
  dataset for land applications}.{\BBCQ}
\newblock
\APACjournalVolNumPages{Earth System Science Data}{13}{9}{4349--4383}.
\PrintBackRefs{\CurrentBib}

\bibitem [\protect \citeauthoryear {%
Pearce%
, Zaki%
, Brintrup%
, Anastassacos%
\BCBL {}\ \BBA {} Neely%
}{%
Pearce%
\ \protect \BOthers {.}}{%
{\protect \APACyear {2018}}%
}]{%
pearce2018uncertainty}
\APACinsertmetastar {%
pearce2018uncertainty}%
\begin{APACrefauthors}%
Pearce, T.%
, Zaki, M.%
, Brintrup, A.%
, Anastassacos, N.%
\BCBL {}\ \BBA {} Neely, A.%
\end{APACrefauthors}%
\unskip\
\newblock
\APACrefYearMonthDay{2018}{}{}.
\newblock
{\BBOQ}\APACrefatitle {Uncertainty in neural networks: Bayesian ensembling}
  {Uncertainty in neural networks: Bayesian ensembling}.{\BBCQ}
\newblock
\APACjournalVolNumPages{stat}{1050}{}{12}.
\PrintBackRefs{\CurrentBib}

\bibitem [\protect \citeauthoryear {%
Pincus%
, Batstone%
, Hofmann%
, Taylor%
\BCBL {}\ \BBA {} Glecker%
}{%
Pincus%
\ \protect \BOthers {.}}{%
{\protect \APACyear {2008}}%
}]{%
pincus2008evaluating}
\APACinsertmetastar {%
pincus2008evaluating}%
\begin{APACrefauthors}%
Pincus, R.%
, Batstone, C\BPBI P.%
, Hofmann, R\BPBI J\BPBI P.%
, Taylor, K\BPBI E.%
\BCBL {}\ \BBA {} Glecker, P\BPBI J.%
\end{APACrefauthors}%
\unskip\
\newblock
\APACrefYearMonthDay{2008}{}{}.
\newblock
{\BBOQ}\APACrefatitle {Evaluating the present-day simulation of clouds,
  precipitation, and radiation in climate models} {Evaluating the present-day
  simulation of clouds, precipitation, and radiation in climate models}.{\BBCQ}
\newblock
\APACjournalVolNumPages{Journal of Geophysical Research:
  Atmospheres}{113}{D14}{}.
\PrintBackRefs{\CurrentBib}

\bibitem [\protect \citeauthoryear {%
Que%
, Ma%
, Ma%
\BCBL {}\ \BBA {} Chen%
}{%
Que%
\ \protect \BOthers {.}}{%
{\protect \APACyear {2020}}%
}]{%
que2020spatiotemporal}
\APACinsertmetastar {%
que2020spatiotemporal}%
\begin{APACrefauthors}%
Que, X.%
, Ma, X.%
, Ma, C.%
\BCBL {}\ \BBA {} Chen, Q.%
\end{APACrefauthors}%
\unskip\
\newblock
\APACrefYearMonthDay{2020}{}{}.
\newblock
{\BBOQ}\APACrefatitle {A spatiotemporal weighted regression model (STWR v1. 0)
  for analyzing local nonstationarity in space and time} {A spatiotemporal
  weighted regression model (stwr v1. 0) for analyzing local nonstationarity in
  space and time}.{\BBCQ}
\newblock
\APACjournalVolNumPages{Geoscientific Model Development}{13}{12}{6149--6164}.
\PrintBackRefs{\CurrentBib}

\bibitem [\protect \citeauthoryear {%
R{\"a}is{\"a}nen%
, Ruokolainen%
\BCBL {}\ \BBA {} Ylh{\"a}isi%
}{%
R{\"a}is{\"a}nen%
\ \protect \BOthers {.}}{%
{\protect \APACyear {2010}}%
}]{%
raisanen2010weighting}
\APACinsertmetastar {%
raisanen2010weighting}%
\begin{APACrefauthors}%
R{\"a}is{\"a}nen, J.%
, Ruokolainen, L.%
\BCBL {}\ \BBA {} Ylh{\"a}isi, J.%
\end{APACrefauthors}%
\unskip\
\newblock
\APACrefYearMonthDay{2010}{}{}.
\newblock
{\BBOQ}\APACrefatitle {Weighting of model results for improving best estimates
  of climate change} {Weighting of model results for improving best estimates
  of climate change}.{\BBCQ}
\newblock
\APACjournalVolNumPages{Climate dynamics}{35}{2}{407--422}.
\PrintBackRefs{\CurrentBib}

\bibitem [\protect \citeauthoryear {%
Sanderson%
, Knutti%
\BCBL {}\ \BBA {} Caldwell%
}{%
Sanderson%
\ \protect \BOthers {.}}{%
{\protect \APACyear {2015}}%
}]{%
sanderson2015representative}
\APACinsertmetastar {%
sanderson2015representative}%
\begin{APACrefauthors}%
Sanderson, B\BPBI M.%
, Knutti, R.%
\BCBL {}\ \BBA {} Caldwell, P.%
\end{APACrefauthors}%
\unskip\
\newblock
\APACrefYearMonthDay{2015}{}{}.
\newblock
{\BBOQ}\APACrefatitle {A representative democracy to reduce interdependency in
  a multimodel ensemble} {A representative democracy to reduce interdependency
  in a multimodel ensemble}.{\BBCQ}
\newblock
\APACjournalVolNumPages{Journal of Climate}{28}{13}{5171--5194}.
\PrintBackRefs{\CurrentBib}

\bibitem [\protect \citeauthoryear {%
Sanderson%
, Wehner%
\BCBL {}\ \BBA {} Knutti%
}{%
Sanderson%
\ \protect \BOthers {.}}{%
{\protect \APACyear {2017}}%
}]{%
sanderson2017skill}
\APACinsertmetastar {%
sanderson2017skill}%
\begin{APACrefauthors}%
Sanderson, B\BPBI M.%
, Wehner, M.%
\BCBL {}\ \BBA {} Knutti, R.%
\end{APACrefauthors}%
\unskip\
\newblock
\APACrefYearMonthDay{2017}{}{}.
\newblock
{\BBOQ}\APACrefatitle {Skill and independence weighting for multi-model
  assessments} {Skill and independence weighting for multi-model
  assessments}.{\BBCQ}
\newblock
\APACjournalVolNumPages{Geoscientific Model Development}{10}{6}{2379--2395}.
\PrintBackRefs{\CurrentBib}

\bibitem [\protect \citeauthoryear {%
Stegall%
\ \BBA {} Kunkel%
}{%
Stegall%
\ \BBA {} Kunkel%
}{%
{\protect \APACyear {2019}}%
}]{%
stegall2019}
\APACinsertmetastar {%
stegall2019}%
\begin{APACrefauthors}%
Stegall, S\BPBI T.%
\BCBT {}\ \BBA {} Kunkel, K\BPBI E.%
\end{APACrefauthors}%
\unskip\
\newblock
\APACrefYearMonthDay{2019}{}{}.
\newblock
{\BBOQ}\APACrefatitle {Simulation of daily extreme precipitation over the
  United States in the CMIP5 30-Yr decadal prediction experiment} {Simulation
  of daily extreme precipitation over the united states in the cmip5 30-yr
  decadal prediction experiment}.{\BBCQ}
\newblock
\APACjournalVolNumPages{Journal of Applied Meteorology and
  Climatology}{58}{4}{875--886}.
\PrintBackRefs{\CurrentBib}

\bibitem [\protect \citeauthoryear {%
Taylor%
, Stouffer%
\BCBL {}\ \BBA {} Meehl%
}{%
Taylor%
\ \protect \BOthers {.}}{%
{\protect \APACyear {2012}}%
}]{%
taylor2012overview}
\APACinsertmetastar {%
taylor2012overview}%
\begin{APACrefauthors}%
Taylor, K\BPBI E.%
, Stouffer, R\BPBI J.%
\BCBL {}\ \BBA {} Meehl, G\BPBI A.%
\end{APACrefauthors}%
\unskip\
\newblock
\APACrefYearMonthDay{2012}{}{}.
\newblock
{\BBOQ}\APACrefatitle {An overview of CMIP5 and the experiment design} {An
  overview of cmip5 and the experiment design}.{\BBCQ}
\newblock
\APACjournalVolNumPages{Bulletin of the American meteorological
  Society}{93}{4}{485--498}.
\PrintBackRefs{\CurrentBib}

\bibitem [\protect \citeauthoryear {%
Ukkola%
, De~Kauwe%
, Roderick%
, Abramowitz%
\BCBL {}\ \BBA {} Pitman%
}{%
Ukkola%
\ \protect \BOthers {.}}{%
{\protect \APACyear {2020}}%
}]{%
ukkola2020robust}
\APACinsertmetastar {%
ukkola2020robust}%
\begin{APACrefauthors}%
Ukkola, A\BPBI M.%
, De~Kauwe, M\BPBI G.%
, Roderick, M\BPBI L.%
, Abramowitz, G.%
\BCBL {}\ \BBA {} Pitman, A\BPBI J.%
\end{APACrefauthors}%
\unskip\
\newblock
\APACrefYearMonthDay{2020}{}{}.
\newblock
{\BBOQ}\APACrefatitle {Robust future changes in meteorological drought in CMIP6
  projections despite uncertainty in precipitation} {Robust future changes in
  meteorological drought in cmip6 projections despite uncertainty in
  precipitation}.{\BBCQ}
\newblock
\APACjournalVolNumPages{Geophysical Research Letters}{47}{11}{e2020GL087820}.
\PrintBackRefs{\CurrentBib}

\bibitem [\protect \citeauthoryear {%
Weigel%
\ \protect \BOthers {.}}{%
Weigel%
\ \protect \BOthers {.}}{%
{\protect \APACyear {2021}}%
}]{%
weigel2021earth}
\APACinsertmetastar {%
weigel2021earth}%
\begin{APACrefauthors}%
Weigel, K.%
, Bock, L.%
, Gier, B\BPBI K.%
, Lauer, A.%
, Righi, M.%
, Schlund, M.%
\BDBL {}others%
\end{APACrefauthors}%
\unskip\
\newblock
\APACrefYearMonthDay{2021}{}{}.
\newblock
{\BBOQ}\APACrefatitle {Earth System Model Evaluation Tool (ESMValTool) v2.
  0--diagnostics for extreme events, regional and impact evaluation, and
  analysis of Earth system models in CMIP} {Earth system model evaluation tool
  (esmvaltool) v2. 0--diagnostics for extreme events, regional and impact
  evaluation, and analysis of earth system models in cmip}.{\BBCQ}
\newblock
\APACjournalVolNumPages{Geoscientific Model Development}{14}{6}{3159--3184}.
\PrintBackRefs{\CurrentBib}

\bibitem [\protect \citeauthoryear {%
Wenzel%
, Eyring%
, Gerber%
\BCBL {}\ \BBA {} Karpechko%
}{%
Wenzel%
\ \protect \BOthers {.}}{%
{\protect \APACyear {2016}}%
}]{%
wenzel2016constraining}
\APACinsertmetastar {%
wenzel2016constraining}%
\begin{APACrefauthors}%
Wenzel, S.%
, Eyring, V.%
, Gerber, E\BPBI P.%
\BCBL {}\ \BBA {} Karpechko, A\BPBI Y.%
\end{APACrefauthors}%
\unskip\
\newblock
\APACrefYearMonthDay{2016}{}{}.
\newblock
{\BBOQ}\APACrefatitle {Constraining future summer austral jet stream positions
  in the CMIP5 ensemble by process-oriented multiple diagnostic regression}
  {Constraining future summer austral jet stream positions in the cmip5
  ensemble by process-oriented multiple diagnostic regression}.{\BBCQ}
\newblock
\APACjournalVolNumPages{Journal of Climate}{29}{2}{673--687}.
\PrintBackRefs{\CurrentBib}

\bibitem [\protect \citeauthoryear {%
Zelazowski%
, Huntingford%
, Mercado%
\BCBL {}\ \BBA {} Schaller%
}{%
Zelazowski%
\ \protect \BOthers {.}}{%
{\protect \APACyear {2018}}%
}]{%
zelazowski2018climate}
\APACinsertmetastar {%
zelazowski2018climate}%
\begin{APACrefauthors}%
Zelazowski, P.%
, Huntingford, C.%
, Mercado, L\BPBI M.%
\BCBL {}\ \BBA {} Schaller, N.%
\end{APACrefauthors}%
\unskip\
\newblock
\APACrefYearMonthDay{2018}{}{}.
\newblock
{\BBOQ}\APACrefatitle {Climate pattern-scaling set for an ensemble of 22
  GCMs--adding uncertainty to the IMOGEN version 2.0 impact system} {Climate
  pattern-scaling set for an ensemble of 22 gcms--adding uncertainty to the
  imogen version 2.0 impact system}.{\BBCQ}
\newblock
\APACjournalVolNumPages{Geoscientific Model Development}{11}{2}{541--560}.
\PrintBackRefs{\CurrentBib}

\end{thebibliography}

\end{document}